\documentclass[12pt, letterpaper]{article}              
\usepackage[includeheadfoot,  
            marginratio={1:1,2:3},  
            width=412pt,  
            height=688pt,]{geometry} 

\usepackage{hypbmsec}

\usepackage{afterpage}
\usepackage[isu,bf]{caption}

\usepackage{annulus} 
\usepackage{amsmath} 
\usepackage{amsfonts} 
\usepackage{amssymb} 
\usepackage{bbm} 
 
\usepackage{graphicx} 
\usepackage{subfigure}

\usepackage{epsfig}
\usepackage{psfig}

\author{}
\newcommand{\drawsquare}[2]{\hbox{%
\rule{#2pt}{#1pt}\hskip-#2pt
\rule{#1pt}{#2pt}\hskip-#1pt
\rule[#1pt]{#1pt}{#2pt}}\rule[#1pt]{#2pt}{#2pt}\hskip-#2pt
\rule{#2pt}{#1pt}}

\newcommand{\fund}{\raisebox{-.5pt}{\drawsquare{6.5}{0.4}}}
\newcommand{\antifund}{\overline{\fund}}

\newcommand{\be}{\begin{equation}}
\newcommand{\ee}{\end{equation}}
\newcommand{\bea}{\begin{eqnarray}}
\newcommand{\eea}{\end{eqnarray}}

\newcommand{\ov}{\overline} 

\def\IR{\relax{\rm I\kern-.18em R}}

\begin{document}

\title{
\begin{flushright} \vspace{-2cm}
{\small MPP-2009-15 \\
UPR-1205-T\\
SLAC-PUB-13531 \\
 \vspace{-0.35cm}
} \end{flushright} \vspace{3cm} D-brane Instantons in Type II String Theory}
\author{}
\date{}

\maketitle

\begin{center}
Ralph Blumenhagen$^1$, Mirjam Cveti\v c$^2$, Shamit Kachru$^{3,4}$ and  Timo Weigand$^4$\\
\vspace{0.5cm}
\emph{$^1$ Max-Planck-Institut f\"ur Physik, F\"ohringer Ring 6, \\
D-80805 M\"unchen, Germany}\\
\vspace{0.1cm} \emph{$^2$ Department of Physics and Astronomy, University of
Pennsylvania,
 Philadelphia, PA 19104-6396, USA }\\
\vspace{0.1cm} \emph{$^3$ Department of Physics, Stanford University, \\
Stanford, CA 94305, USA} \\
\vspace{0.1cm} \emph{$^4$ SLAC National Accelerator Laboratory, Stanford University, \\
Menlo Park, CA 94309, USA}

\vspace{2cm}
\end{center}

\begin{abstract}
\noindent We review recent progress in determining the effects of  
D-brane instantons  in ${\cal N}=1$ supersymmetric compactifications of
Type II string theory to four dimensions.
We describe  the abstract D-brane instanton calculus for holomorphic
couplings such as the superpotential, the gauge kinetic function and higher fermionic F-terms. This includes a discussion of multi-instanton effects and the implications of background fluxes for the instanton sector.
Our presentation also highlights, but is not restricted to the computation of D-brane instanton effects in quiver gauge theories on D-branes at singularities.
We then summarize the 
 concrete consequences of stringy D-brane instantons
 for the construction of semi-realistic models of
particle physics or SUSY-breaking  
in compact and non-compact geometries. 

\end{abstract}

\vfill
\hrulefill\hspace*{4in}


\thispagestyle{empty} \clearpage

\tableofcontents

\section{INTRODUCTION}
\label{sintro}

The main object of interest in any quantum field theory 
with a  perturbative expansion is the computation of correlation functions. In general these correlation functions
are already non-vanishing at tree-level and receive 
perturbative corrections at each loop level. If the relevant
coupling constant $g$  is small, higher loop levels are
suppressed  by powers of $g$. On top of this perturbative series,
non-perturbative corrections can also arise. In the semi-classical
approximation these are associated with topologically non-trivial solutions
to the classical equations of motion \cite{original,Belavin:1975fg}, and contribute terms that scale like
$\exp(-1/g^2)$ to the correlation functions.
Therefore, they are more strongly suppressed than any perturbative
correction.

These a priori subleading non-perturbative corrections can however become very
important when all potentially larger corrections are known to be absent due to non-renormalization
theorems.
Such situations can be realized in  supersymmetric 
quantum field theories. For instance, in the context of ${\cal N}=1$ supersymmetric
four-dimensional theories, there exist holomorphic quantities such as
the superpotential $W$, 
\bea
           S_W=\int d^4 x\, d^2\theta\  W(\phi_i),
\eea
and the gauge kinetic function $f$,
\bea
               S_{\rm Gauge}=\int d^4 x\, d^2 \theta \,\,
              f(\phi_i)\,\, {\rm tr}\left(  W^\alpha\, W_{\alpha}\right)\; ,
\eea
which are only
integrated over half of the superspace and depend holomorphically
on the chiral superfields $\phi_i$. 
At the perturbative level, the superpotential and gauge kinetic function respectively receive only tree-level and 
up to one-loop level contributions \cite{Grisaru:1979wc,Seiberg:1993vc}.
As a consequence, non-perturbative corrections can become very important for the dynamics of
the system, 
in particular if for instance the tree-level superpotential coupling 
vanishes. Since these non-perturbative contributions are
exponentially suppressed in the weak-coupling regime, 
when they are the leading effect they may provide a dynamical
explanation of some of
the hierarchy problems of fundamental physics.

In gauge theories such non-perturbative corrections arise from so-called
gauge instantons. These are solutions to the Euclidean self-duality
equation
\bea
                    F=* F
\eea
for the Yang-Mills gauge field. Such solutions can be explicitly constructed
as local minima of the action and are classified by the instanton number 
$N=\int_{\IR^4}  {\rm tr}F\wedge F$. Around each instanton saddle
point, one can again perform perturbation theory and compute 
the contributions to certain correlation functions. The final result will then involve summation over all topologically non-trivial sectors.
The prescription to carry out these computations is determined by the so-called
instanton calculus. As a main ingredient
it involves integration over the collective coordinates, also known as the moduli space of the instanton solution.

In string theory the situation is very similar. Also here
one can compute perturbative corrections to tree-level correlation
functions.\footnote{Here we are speaking loosely.  In asymptotically AdS solutions, one is computing
correlation functions (of a dual field theory).  In asymptotically Minkowski backgrounds, one computes
an S-matrix, and infers an effective action indirectly.  Then the corrections we discuss are really
to terms in this effective action.  In a non-gravitational theory, this action would give rise to 
meaningful off-shell correlation functions.}
 These are given by two-dimensional 
conformal field theory correlation
functions on Riemannian surfaces of higher genus $g$. The
expansion parameter is $g_s^{(2g-2)}$ with $g_s$  the string coupling and depends on the dilaton $\varphi$ 
via $g_s=\exp(\varphi)$. Suppose we compactify
the ten-dimensional superstring on a six-dimensional background such that
${\cal N}=1$ supersymmetry is preserved in four dimensions. One can then
compute an effective four-dimensional supergravity action  for
the massless modes. The non-renormalization theorems for holomorphic
couplings generalize naturally to the string case. 
For the holomorphic couplings $W$ and $f$ we expect that beyond
tree- and one-loop level, respectively,  non-perturbative corrections are present.
Since we still lack a complete second quantized version of string
theory, one must argue for the existence of these non-perturbative corrections by
the analogy with field theory.  Stringent tests of their presence in decoupling limits,
or in cases where duality maps such effects to classical effects, provide overwhelming
evidence
that this analogy is correct.

From the early days of the heterotic string, effects that are non-perturbative from the worldsheet perspective have been a  field of active interest. Such configurations arise as Euclidean closed
strings wrapping topologially non-trivial two-cycles of the compactification manifold \cite{dsww86,dsww87}. Being localized in four dimensions, they are called worldsheet instantons, in analogy to the Euclidean topologically non-trivial solutions of Yang-Mills theory. Their contribution to the couplings is non-perturbative
in the worldsheet expansion parameter $\alpha'$, but not in the string coupling $g_s$.

However, the past one and a half decades have witnessed major progress
in the understanding of  objects in string theory which are non-perturbative also from the spacetime point of view.
It has been shown that p-brane solutions of the supergravity
equations of motions are truly non-perturbative objects in
string theory. In particular for the large class of D-branes,
the quantum theory around the classical solution
is known to be  given by an open string theory with
end-points on the D-brane \cite{JP95}. These D-branes carry charge
under certain Ramond-Ramond p-forms and also have tension
scaling like $T_p=g_s^{-1}$. 
Such objects are indeed present in four-dimensional Type II string vacua preserving ${\cal N}=1$ space-time supersymmetry in two different ways.
Firstly, D-branes can fill four-dimensional space-time and  wrap
certain cycles of the internal manifold. These D-branes
carry both gauge fields and chiral
matter fields which are observed as physical fields on the four-dimensional effective theory localized on the D-brane. The past years have seen 
many attempts to realize realistic gauge and matter spectra
on such intersecting D-brane models. This includes 
investigations of D-branes on compact manifolds as
well as on non-compact geometries. Beyond the mere construction of models,
a formalism has been developed to compute the resulting
${\cal N}=1$ supersymmetric four-dimensional effective supergravity
action. 

Soon it was realized that D-branes play an important role not only as the  hosts of this effective field theory, but also as actors in it:
 Euclidean D-branes wrapping entirely a topologically non-trivial cycle of the internal manifold appear as  truly pointlike objects in space and time and thus deserve the name D-brane instanton \cite{EW95a,EW95b,MD95}. In addition, closed and open \cite{kklm99} worldsheet instantons lead to corrections which are non-perturbative in the string tension $\alpha'$, just as in the heterotic cousin theory. 
The analogy with Yang-Mills theory can be made very explicit also for D-brane instantons: There the groundstate, or vacuum, of \texttt{the the}ory is given by a trivial field configuration, and in computing correlation functions one has to sum over all non-trivial configurations, each of which is described in a perturbative saddle-point approximation. Here the vacuum corresponds to a stable configuration of spacetime-filling D-branes, and the topologically non-trivial situations include these Euclidean Dp-branes, or Ep-branes in short. 
Open string perturbation theory can be used to describe the fluctuations of Ep-branes and an instanton calculus can be defined in analogy to the field theory case \cite{JP94,bbs95,hm99,gg97a,gg97b,MB02}.

In special situations these D-brane instantons reproduce the ADHM construction of gauge instantons in the field theory limit. 
But D-brane instantons are not restricted to a microscopic realization of gauge instanton effects. Rather, they can generate superpotential contributions independent of the gauge degrees of freedom, as pioneered in \cite{EW96}. For example the non-perturbative generation of a superpotential for some of the closed string fields is crucial in attempts to stabilize the massless moduli fields of four-dimensional compactifications \cite{kklt03}. The reason why non-perturbaive effects become important here is precisely the absence of competing terms at the perturbative level which are forbidden due to the non-renormalization of the superpotential.

More recently it has become clear that D-brane instantons play a crucial role for the same reason also in the open string sector of intersecting brane worlds.
Oftentimes the presence of global $U(1)$ symmetries
forbids some of the phenomenologically desirable matter
couplings such as Majorana neutrino masses, Yukawa couplings  or the $\mu$-term.
It was found that D-brane instantons can contribute
to these quantities by effectively breaking the global symmetries \cite{bcw06,iu06,fkms06} (see also \cite{hklvz06}).
The exponential suppression by the classical instanton action ${\rm exp}{(- {\rm Vol_{E_p}}/{g_s})}$ depends on the volume (in string units) of the cycle wrapped by the instanton. As such it is in general independent of the gauge coupling in four dimensions. This property, which resulted in the name \emph{stringy} or \emph{exotic instantons}, explains why the non-perturbatively generated couplings can become relevant even in  situations where $g_s$ is perturbatively small.
There are also obvious potential hidden sector applications of such stringy instantons.  For instance, they can lead to
models of dynamical supersymmetry breaking without strongly coupled field theory dynamics, and with very minimal
hidden sectors \cite{Aharony:2007db}.  
These and other applications of stringy instantons have been intensely explored in the past years, and their
relevance for the physics of string compactifications has revived interest also in more technical aspects of instanton calculus.

The aim of this review is both to give a pedagogical introduction
to recent developments in the study of D-brane instanton effects
in Type II string theory, and to provide an overview of the
various generalizations and applications which have appeared
during the last couple of years.
Due to lack of space we have to assume some knowledge
of four-dimensional ${\cal N}=1$ supersymmetric
Type II orientifold compactifications. These have been
an active field of research in the recent past and
a number of review articles exist including \cite{as02,bcls05,dk06,bkls06,AU07,FM07,FD08,DL08}.
For the local string models, a certain familiarity with
D-branes at singularities and the resulting quiver
gauge theories is also helpful; nice reviews appear in \cite{Malyshev:2007zz,Wijnholt:2007vn}. Some aspects of D-brane instantons are also covered in the recent review articles \cite{abls07b,crw07b,NA08,MB08c}.

We will begin in \S\ref{sBasics} with a general classification of D-brane instantons in Type II orientifold models. 
The precise couplings a given D-brane instanton can generate are determined by the zero mode content of its worldvolume theory. We then outline the rules for the computation of an instanton induced superpotential. In \S\ref{sGenerationFterms} we classify which quantities of the four-dimensional effective action are corrected by D-brane instantons of different kinds, where for the sake of brevity we have to focus on holomorphic objects. After reviewing the special case of gauge instanton effects we discuss corrections to the gauge kinetic function and higher fermionic F-terms. Consistency of the instanton calculus automatically requires the inclusion also of multi-instanton effects. Closed string background fluxes, which play a crucial role in the stabilization of massless moduli fields, modify the details of all these effects by lifting some of the fermionic zero modes. We conclude this technical section with a brief summary of known instanton contributions to D-terms.  In \S\ref{secquiver}, we describe how one can apply the D-instanton calculus, most easily
derived in the class of free worldsheet conformal field theories, to situations which involve non-trivial geometries, e.g. branes at
singularities.  We find that the rules generalize in a straightforward manner, with the interactions an instanton can generate
being determined entirely by data which is present in the quiver gauge theory of spacetime filling branes at the same singularity.
We also describe how the powerful techniques of geometric transitions can be used, for some of these cases, to provide
a dual computation of the instanton-generated superpotential, including the precise coefficients and multi-cover
contributions to the superpotential.   This provides a highly non-trivial check on our considerations.  Finally, we 
describe some relations between stringy instantons and conventional Yang-Mills instantons in cascading gauge
theories, that provide an alternative
check on the stringy results.
In \S\ref{sphenoappl} we very briefly discuss various possible phenomenological applications of these abstract considerations,
and we close with a discussion of future directions in \S\ref{sconclusions}.

\section{BASICS ON BPS D-BRANE INSTANTONS}
\label{sBasics}

\subsection{Classification of D-brane instantons}
\label{subsecDinst}

Let us compactify ten-dimensional Type II string theory on a Calabi-Yau manifold
${\cal X}$ to four-dimensional Minkowski space $\IR^{1,3}$. This
preserves eight supercharges corresponding to ${\cal N}=2$ 
supersymmetry in four dimensions. 
To break the ${\cal N}=2$ space-time supersymmetry down to a 
phenomenologically more appealing  ${\cal N}=1$ supersymmetry, 
one performs an orientifold projection $\Omega_p$.
We now  summarize some of the main features of such orientifold models. 
For more details 
 we refer the reader to the review articles  \cite{bcls05,dk06,bkls06}. 
One distinguishes three kinds of orientifolds models with
very similar features. 

\subsubsection*{Type I models}

The starting point is the Type IIB superstring in ten dimensions.
Taking the quotient by the worldsheet parity symmetry $\Omega: (\sigma,\tau)\to
(-\sigma,\tau)$ one obtains
the well-known Type I string. Since the orientifold acts trivially on the spacetime coordinates the theory exhibits an $O9$-plane. 
The resulting tadpole can be canceled by introducing
stacks of D9-branes carrying generically non-trivial 
stable vector bundles. 
Moreover, there can be D5-branes wrapping holomorphic curves
of the background Calabi-Yau geometry.

The four-dimensional holomorphic superpotential
receives  perturbative contributions only at tree-level
and  depends solely on the
complex structure closed string moduli, i.e.  $W_0(U_i)$.
In such string models, the abelian gauge anomalies are canceled
by a generalized Green-Schwarz mechanism, in which the shift symmetry of
the axions related to the RR-forms $C_2$ and $C_6$ is gauged.
The shift symmetries of axions are generally violated by instantons.
In the present case the relevant objects are the ones coupling to
$C_2$ and $C_6$. These are Euclidean E1- and E5-branes wrapping
internal two-cycles of the Calabi-Yau or the whole Calabi-Yau, respectively.
The classical instanton action is given by the volume of these internal cycles,
which are complexified into chiral multiplets as
\bea
\label{superfieldsa}
          T_i=e^{-\varphi} \int_{\Gamma^i_2} J +i \int_{\Gamma^i_2} C_2, \qquad
          S=e^{-\varphi}   \int_{\cal X} J^3 +i \int_{\cal X} C_6.
\eea
One therefore expects that  the superpotential can receive
contributions from E1-and E5-instantons and has the following
schematic form
\bea 
\label{supernon} 
       W=W_{0}(U)  + \sum_{E1_a}   A_a(U)\,  e^{- \alpha^i_a T_i } + A_S (U)\,  e^{-S}.
\eea 
Similarly, the holomorphic gauge kinetic function on a stack
of D9-branes must look like 
\bea 
\label{gaugenonB} 
    f_A=S - \sum_i \kappa^i_A   T_i + f_A^{\rm{1-loop}}\left( U \right)
       + \sum_{E1_a}   A_a(U)\,  e^{- \alpha^i_a T_i } + A_S (U)\,  e^{-S}.
\eea 
Here $\Gamma_i$ denotes a basis of $H_4(X,\mathbb Z)$ and 
$\kappa^i_A=\int_{\Gamma_i} {\rm ch}_2(V_A)$ depends on the second Chern character of the vector bundle $V_A$ defined on the D9-branes stack with gauge kinetic function $f_A$.
Note that its one-loop correction must not depend 
on the K\"ahler  moduli.

\subsubsection*{Type IIB orientifolds with O7 and O3-planes}

One can generalize the orientifold projection by dressing
$\Omega$ with a holomorphic involution $\sigma$. In case
$\sigma$ acts like
\bea
      \sigma: J\to J, \qquad \Omega_3\to -\Omega_3
\eea
the fixed-point set consists of $O7$-planes wrapping holomorphic
four-cycles of the Calabi-Yau and a number of $O3$-planes
localized at certain fixed points of $\sigma$ on ${\cal X}$.
The axion whose shift symmetry is gauged by the Green-Schwarz
mechanism is the RR-form  $C_4$. One thus expects E3-branes wrapping four-cycles in the
Calabi-Yau to contribute to the holomorphic couplings.
In addition there could also be E(-1) brane instantons.
The complexified K\"ahler moduli and the axio-dilaton field read 
\bea 
\label{superfieldsb}
     T_i= e^{-\varphi} \int_{\Gamma^i_4} J\wedge J + i \int_{\Gamma^i_4} C_4,
   \qquad  \tau= e^{-\varphi} +i C_0.
\eea 
The holomorphic quantities can then have an expansion of the form
\bea 
\label{supernonB} 
       W=W_{0}(e^{-T_i})  + \sum_{E3_a}   A_a(U )\,\,  e^{- \alpha^i_a T_i } 
              +  A_{\tau}(U) \,  e^{- \tau } 
\eea 
for the superpotential and 
\bea 
\label{gaugenon} 
    f_A=\sum_i \kappa_A^i \,  T_i + f_A^{\rm{1-loop}}\left(U \right) 
     + \sum_{E3_a}   A_a(U)\,\,  e^{- \alpha^i_a T_i }  +  A_{\tau}(U) \,  e^{- \tau }   
\eea  
for the gauge coupling on a stack of D7-branes.

\subsubsection*{Intersecting D6-brane models}

Here one starts with the Type IIA superstring compactified on a
Calabi-Yau and takes the quotient by the orientifold
projection $\Omega\ov\sigma (-1)^{F_L}$, where $\ov\sigma$ denotes
an isometric anti-holomorphic involution.
The fixed-point locus of such an involution is a special Lagrangian
three-cycle in ${\cal X}$ which gives rise to $O6$-planes. Their  
tadpole can be canceled by introducing intersecting D6-branes.
The axion whose shift symmetry is gauged by the Green-Schwarz
mechanism is the RR-form  $C_3$ 
so that one expects E2-branes wrapping three-cycles in the
Calabi-Yau to contribute to the holomorphic couplings.
In this case the instanton action depends on the complex
structure moduli
\bea 
\label{superfieldsc}
     U_i= e^{-\varphi} \int_{\Gamma^i_3}  \Omega_3 + i \int_{\Gamma^i_3} C_3.
\eea 
However, also the complexified K\"ahler moduli
$T_i= \int_{\Gamma^i_2} J_2 + i\int_{\Gamma^i_2} B_2$ contain
as the imaginary part an axion. In combination these observations allow  an  expansion 
\bea 
\label{supernonA} 
       W=W_{0}(e^{-T_i})  + \sum_{E2_a}   A_a(e^{-T_i} )\,\,  e^{- \alpha^i_a U_i } 
\eea 
and
\bea 
\label{gaugenonA} 
    f_A=\sum_i \kappa_A^i \,  U_i + f_A^{\rm{1-loop}}\left(e^{-T_i}\right) 
     + \sum_{E2_a}   A_a(e^{-T_i} )\,\,  e^{- \alpha^i_a U_i }  
\eea  
for the holomorphic quantities.

\subsubsection*{Unified description of all orientifold models}

It is obvious form the above that all three kinds of orientifold models
are very similar in structure. This is a consequence of
T-dualities (mirror symmetry) connecting the three kinds
of orientifolds.
In the sequel we will treat all orientifold models
on the same footing by introducing a unified notation.

Let us denote  the space-time filling D-branes wrapping internal
cycles $\Gamma_a$ as ${\cal D}_a$. For branes in Type IIB these objects 
also carry  non-trivial holomorphic vector bundles ${\cal V}_a$.
Moreover, these objects are not invariant under
the orientifold projection $\Omega_p$ and are mapped
to space-time filling D-branes $({\cal D}'_a, {\cal V}'_a)$.
The corresponding D-brane instantons are denoted as ${\cal E}_i$
and ${\cal E}'_i$. 
At the intersection of two D-branes ${\cal D}_a$ and ${\cal D}_b$ 
one gets matter fields  $\Phi_{a,b}$. Our convention is that
an open string stretching from ${\cal D}_a$ to ${\cal D}_b$ yields a matter $\Phi_{a,b}$ in the bifundamental representation $(\ov N_a, N_b)$. Their multiplicity 
can be computed by the relevant topological cohomology groups.
For $D6$-branes in Type IIA these are simply the positive
and negative intersections between the two 3-cycles. 
For all three cases we use $I^+_{ab}$ to denote positive chirality
fields and  $I^-_{ab}$ for negative chirality
in the $a \rightarrow b$ sector. 
A positive  chiral index $I_{ab}= I^+_{ab} - I^-_{ab}$ then indicates an excess of chiral fields in the representation $(\ov N_a, N_b)$ over those transforming as $(N_a, \ov N_b)$.  For Type IIA orientifolds $I_{ab}$ is simply
the topological intersection number between the two
internal 3-cycles.  
In Type IIB the chiral index is given by a   unified formula in terms
of the K-theoretic intersection number
\bea
      I_{ab}= I^+_{ab}-I^-_{ab}= 
        \int_{\cal X} Q({\cal D}_a,{\cal V}_a) \wedge
                      Q({\cal D}_b,{\cal V}^\vee_b)
\eea
with
\bea
      Q({\cal D}_a,{\cal V}_a)=[\Gamma_a]\wedge {\rm ch}({\cal V}_a)\wedge
         \sqrt{{ {\hat A}(T_{\Gamma_a}) \over  {\hat A}(N_{\Gamma_a}) }}\; .
\eea
Here $[\Gamma_a]$ denotes the Poincar\'e dual of the cycle $\Gamma$ the
D-brane is wrapping and $T_{\Gamma_a}$ and $N_{\Gamma_a}$ are its
tangent and normal bundle.

\subsection{Zero modes}
\label{subsec_zero}

In the previous section we have classified which types of instantons might in principle
correct the holomorphic quantities of the effective action. Our arguments were only based on non-renormalisation theorems and knowledge of the chiral superfields. The actual computation of these corrections requires precise control over the instanton zero modes. These are the massless excitations of open strings with both ends on the same instanton or at the intersection
between two instantonic E-branes or between one D-brane and one
E-brane.  As such they can be computed with standard open string CFT methods and one can associate vertex operators with them.
The only difference with respect to the more familiar case of massless modes between spacetimes filling D-branes results from the four Dirichlet-Neumann conditions of the instanton in the extended four dimensions. In particular, one cannot attribute four-dimensional momentum to the instanton modes so that only massless modes can be considered as on-shell states.
With this caveat in mind one can formally compute couplings between the instanton modes among themselves and possibly involving some of the open strings in the D-brane sector.

Let us denote the collection of all instanton zero modes as ${\cal M}$.
The zero mode couplings are encoded in the interaction part of the instanton effective action $S_{\cal E}^{(\rm int.)}({\cal M})$. In  analogy with standard wisdom for field theory instantons, the  instanton contribution to the four-dimensional effective action is sketchily given by
\bea
\label{saturation}
S^{4D}_{n.p.} = \int d{\cal M} \, \, e^{- S^{(0)}_{\cal E} - S_{\cal E}^{(\rm int.)}({\cal M})},
\eea
where  $S^{(0)}_{\cal E}$ denotes the classical instanton effective action given by the complexified superfields of the previous section. We will be much more precise in section \ref{subsecsuperpot}.

Of special importance are  the fermionic zero modes. 
The integral $\int d {\cal M}$ over the zero mode measure requires that each fermionic zero modes can be pulled down from the exponent precisely once as otherwise the Grassmannian integral vanishes. This process is called saturation of fermionic zero modes. 
Oftentimes knowledge of the fermionic zero mode content is therefore sufficient to decide whether or not an instanton
can contribute to a certain correlation function.

After these preliminaries we classify the various kinds of zero modes
of an E-brane instanton in Type II orientifolds.

\subsubsection*{Universal zero modes} 
The universal zero modes arise from strings starting and ending on the same E-brane. 
Firstly, there are four bosonic zero modes $x^{\mu}$ parameterizing the
position of the instanton in four-dimensional spacetime. These are the Goldstone bosons associated with the breakdown of four-dimensional translational invariance due to the presence of the instanton.

Secondly, there are fermionic zero modes related to broken supersymmetries.
In general the instantonic brane ${\cal E}$ is not invariant
under the orientifold projection and there exists an image
brane ${\cal E}'$ wrapping a distinct cycle. In this case, the instanton locally feels the 
full ${\cal N}=2$ supersymmetry preserved by compactification of
Type II theory on a Calabi-Yau manifold \cite{abfk07,abflp07,bfm07,isu07}. The orientifold action preserves a specific ${\cal N}=1$  subalgebra thereof with supercharges $Q^{\alpha}, \ov Q^{\dot \alpha}$.  The orthogonal ${\cal N}=1$  complement, generated by the charges  $Q'^{\alpha}, \ov Q'^{\dot \alpha}$, is broken in four dimensions. A spacetime-filling D-brane along a 1/2-BPS cycle which is supersymmetric with respect to the orientifold preserves the supercharges $Q^{\alpha}, \ov Q^{\dot \alpha}$.  
An instanton along this same cycle likewise preserves 
four of the eight supersymmetries, leading to
four goldstino fermionic zero modes associated with the four broken supersymmetries \footnote{Note that if the instanton did not preserve any supersymmetry,
i.e. if it were non-BPS, 
one would get eight fermionic zero modes, so that such an object
cannot contribute to any supersymmetric quantity.}. However, due to its localisation in the four extended dimensions the BPS-instanton does not preserve the four supercharges $Q^{\alpha}, \ov Q^{\dot \alpha}$, but rather the off-diagonal combination $Q'^{\alpha}, \ov Q^{\dot \alpha}$. Therefore the two chiral Goldstone modes $\theta^{\alpha}$ associated with the breaking of $Q^{\alpha}$ really correspond to half of the ${\cal N}=1$ superspace preserved in four dimensions, while the anti-chiral Goldstinos, denoted by $\ov \tau^{\dot \alpha}$ for distinction, are associated with its orthogonal complement in the original ${\cal N}=2$ algebra \cite{bcrw07}. 
\begin{table}[ht]
\centering
\begin{tabular}{|c|c|}
\hline
${\cal N}=1$  & ${\cal N}=1'$   \\
\hline \hline
$    \theta^{\alpha}  $ & $ \tau^{\alpha}  $   \\ \hline
$ {  \ov \theta^{\dot \alpha} } $ & $ {\ov \tau^{\dot \alpha} } $   \\
\hline
\end{tabular}
\caption{Universal fermionic zero modes $\theta^{\alpha}, \ov \tau^{\dot \alpha}$ ($\tau^{\alpha}, \ov \theta^{\dot \alpha})$ of an (anti-)instanton associated with the breaking of the ${\cal N}=1$ SUSY algebra preserved by the orientifold and its orthogonal complement ${\cal N}=1'$.
\label{tabGoldstino} } 
\end{table}
It follows that such an 
instanton can in principle contribute to an F-term provided some mechanism is found which saturates the extra $\ov \tau^{\dot \alpha}$ modes. 
We will systematically discuss various possibilities in section \ref{sGenerationFterms}.

It is important to appreciate, though, that the generation of a D-term requires instead the four Goldstone modes $\theta^{\alpha}, \ov \theta^{\dot \alpha}$. We will comment on appropriate configurations in section \ref{sec_D-term}.

The simplest configuration leading to an F-term is given by an instanton
that already locally feels only the
${\cal N}=1$ supersymmetry. Since the breaking of  
${\cal N}=2$ to ${\cal N}=1$ occurs on the orientifold
planes and on the space-time filling D-branes, we must
either place the instanton on top of a D-brane
or place it in an $\Omega_p$ invariant position.

The first case, ${\cal E}\subset {\cal D}$,
is nothing else than the stringy description of a gauge instanton
for the Chan-Paton gauge theory on the D-brane ${\cal D}$.
This can be made very precise in that one can
derive the celebrated ADHM constraints by evaluating open string
disc diagrams and is the subject of \S\ref{sgaugeinst}. As will be detailed there, such instantons contribute to the superpotential even in presence of just a single parallel brane ${\cal D}$ even though they cannot directly be interpreted as gauge instantons of the associated $U(1)$ theory.

For the second case with ${\cal E}={\cal E}'$ one has
to distinguish the possibilities that the $\Omega_p$ projection
either symmetrizes or anti-symmetrizes the CP-gauge group \cite{abfk07,abflp07,bfm07,isu07}.
Special care has to be taken due the four Dirichlet-Neumann (DN) boundary conditions
between ${\cal E}$ and an auxiliary  D-brane ${\cal D}$ wrapping the
same internal cycle.  
If the Chan-Paton gauge group on ${\cal D}$ is
$(S)O(N)/SP(N)$,  then the CP-gauge group on ${\cal E}$ is
switched to $SP(N)/(S)O(N)$. 
This is because the action on the Chan-Paton factors switches from symmetrization to anti-symmetrization and vice versa.
In addition, the orientifold acts with an extra minus sign on chiral spinors as well as on bosonic excitations along the four extended directions. More details can be found in \cite{abfk07,abflp07,bfm07,isu07,bcrw07}.
The result can be phrased as follows:
\begin{itemize}
\item $SP(N)$ instanton: In this case one needs an even number of branes
           ${\cal E}$. As a result one gets 
            ${N(N-1)\over 2}$ zero modes $\theta^\alpha$  and
            ${N(N+1)\over 2}$ zero modes $\ov{\tau}^{\dot\alpha}$.
           For $N=2$ one therefore has one chiral and three anti-chiral Weyl-spinors.

\item $O(N)$ instanton: Here one can also have an odd number of  ${\cal E}$ branes.
          One gets 
            ${N(N+1)\over 2}$ zero modes $\theta^\alpha$  and
            ${N(N-1)\over 2}$ zero modes  $\ov\tau^{\dot\alpha}$.
          For $N=1$ one therefore ends up with two
                universal zero modes $\theta^{\alpha}$.
\end{itemize}

On the open string worldsheet, these universal zero modes  are described by 
the vertex operator (in the $(-1/2)$-ghost picture)
\bea
     V^{-{1\over 2}}_{\theta}(z)=\theta_\alpha\,\, e^{-{\varphi(z)\over 2}}\,\, 
     S^\alpha(z)\,\, \Sigma^{{\cal E},{\cal E}}_{{3\over 8},{3\over 2}}(z),
\eea
where $\theta_\alpha$, $\alpha=1,2$  is the polarization and $S^\alpha$ 
denotes the 4D spin field of $SO(1,3)$. This
is a Weyl spinor  of conformal dimension $h={1/4}$. 
The twist field $\Sigma^{{\cal E},{\cal E}}_{{3\over 8},{3\over 2}}$
is essentially the spectral flow operator of the 
${\cal N}=2$ superconformal field theory. The subscripts denote its conformal dimension $\frac{3}{8}$ and $U(1)$ worldsheet charge  ${3\over 2}$. A pedagogical introduction to the basics of the  ${\cal N}=2$ superconformal field theory can be found e.g. in \cite{BG97}.
Since the instantonic brane ${\cal E} $ wraps a cycle on the CY manifold,
the four-dimensional spacetime is transversal so that there appears  
no four-dimensional momentum factor  in the vertex operator.

Finally let us introduce some nomenclature: 
D-brane instantons along a cycle not populated by a D-brane, which do therefore not directly have a gauge instanton
description, have been called {\it stringy} or {\it exotic} instantons
in the literature. Stringy instantons on top of an orientifold leading to a universal zero mode measure $\int d^4 x d^2 \theta$ are known as $O(1)$ instantons, as opposed to so-called $U(1)$ instantons ${\cal E}$ in a non-invariant position ${\cal E} \neq {\cal E'}$.

\paragraph{Deformation zero modes} There can be further zero modes
from the ${\cal E}-{\cal E}$ sector  
due to possible deformations of the instantonic brane.
For a $U(1)$ instanton, each complex valued deformation
leads to one complex bosonic zero mode as well as one chiral and one anti-chiral Weyl spinor, making a total of four fermionic degrees of freedom.
If the instanton is however of type $O(1)$, then
the orientifold projection acts also on these deformation
zero modes as shown in Table \ref{tabledeform}.

\begin{table}[ht]
\centering
\begin{tabular}{|c||c|c|c|}
\hline
zero modes&  E1  & E3  & E2    \\
\hline \hline
$\gamma^{\alpha}$ & $H^{(1,0)}(E1)$  & $H^{(1,0)}(E3)$  &  $b^1(E2)_-$   \\
\hline
$(c, \ov\chi^{\dot\alpha})$ & $H^{0}(E1, N)$ &   $H^{(2,0)}(E3)$  &  $b^1(E2)_+$\\
\hline
\end{tabular}
\caption{Deformation zero modes for three classes of Type II orientifolds.
\label{tabledeform} } 
\end{table}
There the cohomology classes of type $H^{(1,0)}(E)$ count
the Wilson-line moduli and $H^{0}(E, N)$ the transversal deformations
of the cycle $N$. For E2 instantons $b^1(E2)_\pm$
count the moduli even and odd under the orientifold projection.
As anticipated an instanton with deformation moduli can only
contribute to a correlation function if the fermionic
zero modes can be soaked up or lifted by flux, as will be discussed in  \S\ref{sec_gaugekin}, \S\ref{sec_BW} and \S\ref{sec_flux}. An instanton which does not have
these extra zero modes from the ${\cal E}_i-{\cal E}_i$ sector
is also called {\it rigid}.

\subsubsection*{Charged zero modes} 
Finally, there will generically be zero modes which 
arise at the intersection of the instanton ${\cal E}$ with the D-branes 
${\cal D}_a$ \cite{bcw06,iu06,fkms06}.  These zero modes are called {\it charged zero modes}
 as they are charged under the four-dimensional gauge symmetry localized
on the ${\cal D}_a$ branes. They were first discussed in the context of F-theory compacitifications in \cite{OG96}.
Let us focus here on the case of a stringy instanton with a non-trivial intersection with a D-brane along a different cycles.
Due to the four Neumann-Dirichlet boundary conditions between 
${\cal E}$ and ${\cal D}_a$ along the four space-time directions, 
the zero point energy in the NS-sector is already shifted
by $L_0=1/2$ so that for internally intersecting
branes, there can only be fermionic zero modes from the R-sector. The GSO-projection only allows for \emph{chiral} zero modes from the worldsheet point of view \cite{crw07} corresponding to a single Grassmannian degree of freedom. 
The total number of such charged fermionic zero modes is displayed  
for a $U(1)$ instanton in table \ref{tablezero}.
If the instanton wraps the same cycle as a spacetime filling D-brane $\cal D$ there exist also bosonic modes in the ${\cal E} - {\cal D}$  sector. These will be discussed in \S\ref{sgaugeinst}.

\begin{table}[ht]
\centering
\begin{tabular}{|c|c|c|}
\hline
zero mode&  Reps. & number   \\
\hline \hline
$\lambda_{a} \equiv \lambda_{{\cal E}a}$ &    $(-1_{\cal E},\fund_a)$ & $I^+_{{\cal E},{\cal D}_a}$    \\
$\ov\lambda_{a}\equiv \lambda_{a{\cal E}} $ &  $(1_{\cal E},\antifund_a)$  & $I^-_{{\cal E},{\cal D}_a}$    \\
\hline
$\lambda'_{a}\equiv \lambda_{{\cal E}'a} $ &  $(1_{\cal E},\fund_a)$ & $I^+_{{\cal E}',{\cal D}_a}$    \\
$\ov\lambda'_{a} \equiv \lambda_{a{\cal E}'}$ &  $(-1_{\cal E},\antifund_a)$  & $I^-_{{\cal E}',{\cal D}_a}$    \\
\hline
\end{tabular}
\caption{Charged fermionic zero modes from ${\cal E}-{\cal D}$ intersections.
\label{tablezero} } 
\end{table}
From table \ref{tablezero} it is clear that the total $U(1)_a$ charge of all 
the fermionic zero modes on the intersection of ${\cal E}+{\cal E'}$ and 
${\cal D}_a+{\cal D'}_a$
is $Q_a({\cal E})={N}_a\, 
       \left(I_{{\cal E},{\cal D}_a} -I_{{\cal E},{\cal D}'_a}\right)$.
For the case of an $O(1)$ instanton the table simplifies
as ${\cal E}={\cal E}'$ and only the first two lines in Table \ref{tablezero}
give independent zero modes.

In order for such an instanton to contribute to a coupling,
the charged zero modes have to be soaked up.
This is possible because the part $S_{inst.} ({\cal M})$ in equ. (\ref{saturation}) contains couplings of the schematic form
$\lambda_{{\cal E} a_i}\, \Phi_{a_i b_i}  \lambda_{ b_i {\cal E}}  $. The saturation of the $\lambda$ modes thus
pulls
down charged matter fields $\Phi_{a_i b_i}$. This
happens in such a way that for each absorption diagram
the (global) $U(1)_a$ charges are preserved. Therefore
for the superpotential, only terms like 
\bea
 W = \prod_{i=1}^M \Phi_{a_i b_i}\,  \exp \left( - S_{\cal E} \right)
 \label{superpotex}
\eea
can be generated for which the $U(1)_a$ charges of the 
product of matter fields $\prod_i \Phi_i$ are canceled
by the sum of the $U(1)_a$ charges of the zero modes, i.e.
\bea
\label{instcharge}
   \sum_{i=1}^M Q_a(\Phi_{a_i b_i})= - {N}_a\, 
    \left(I_{{\cal E},{\cal D}_a} -I_{{\cal E},{\cal D}'_a}\right)\; .
\eea
It was shown in \cite{bcw06,iu06} that this relation can also be deduced
by using the gauging of the axionic shift symmetries
of the RR-forms $C_{p}$, $p=2,4,3$  due to the generalized Green-Schwarz
mechanism. 
Therefore, such instantons at non-trivial intersection with
 D-branes can generate charged matter couplings which
per se violate the global $U(1)_a$ symmetries.

Let us also give the worldsheet description of these matter field
zero modes. The corresponding
Ramond sector open string vertex operators are of the form
\bea
          V^{-{1\over 2}}_{\lambda^i_{a}}(z)=\lambda^i_{a}\, 
   e^{-{\varphi(z)\over 2}}\,\, 
      \Sigma^{{\cal D}_a,{\cal E}}_{{3\over 8},-{1\over 2}}(z)\,\,
                             \sigma_{h=1/4}(z),
\eea
Here $\Sigma^{{\cal D}_a,{\cal E}}_{{3\over 8},-{ 1\over 2}}$ 
denotes a spin field with conformal dimension $h= {3\over 8}$ and $U(1)$ worldsheet charge $- {1\over 2}$ in the R-sector for the internal SCFT
and $\sigma_{h=1/4}$  the 4D spin field arising from the
twisted 4D worldsheet bosons carrying half-integer modes.
Note that also these zero modes carry no momentum along the flat 4D
directions.

\subsubsection*{Multi-instanton zero modes}

So far we have focused on the zero modes associated with single instantons, but a general configuration may feature several instantonic branes at once. In this case there appear new modes in the sector between two different instantons. 
In fact, such multi-instanton configurations are almost inevitable in the context of Type II orientifolds. Recall that if an instanton ${\cal E}$ wraps a cycle not invariant under the geometric orientifold action we have to consider in addition its orientifold image ${\cal E}'$. This leads, in the upstairs geometry prior to orientifolding, to a two-instanton configuration consisting of $\cal E$ and $\cal E'$. 

As in the case of spacetime-filling D-branes the intersection locus of two instantons ${\cal E}_1$ and ${\cal E}_2$  hosts massless zero modes in form of one chiral multiplet together with its CPT conjugate.
These are counted by the same intersection numbers as in the corresponding ${\cal D}_1-{\cal D}_2$ case. 
Note that this is in contrast with the charged zero modes in the ${\cal E}-{\cal D}$ sector, where 
the boson is projected out  due to the four DN boundary conditions in the extended spacetime dimensions and only a single Grassmann degree of freedom $\lambda$ survives.

Special care has to be taken, though, of the orientifold action for zero modes between an instanton and its image on top of the orientifold plane. 
As encountered before, the orientifold action on the CP factors changes in the instanton sector due to the localization of the instanton in four dimensions and the orientifold acts with an extra minus sign on the chiral Weyl spinors. 
Together with the non-projected sector away from the orientifold locus this gives rise to the zero mode content for ${\cal E}-{\cal E}'$ instantons displayed in table \ref{antizero} \cite{bcrw07}.
\begin{table}[h]
\centering
\begin{tabular}{|c|c|c|}
\hline
zero mode & $U(1)_{\cal E}$ charge &   Multiplicity \\
\hline \hline
$(m, \ov \mu^{\dot \alpha})$ & $(2, -2)$ & $ \frac{1}{2} \left( I_{{\cal E}',{\cal E}} + p  I_{O,{\cal E}}  \right)^+  $  \\
$\mu^{\alpha}$ & $2$ & $ \frac{1}{2} \left( I_{{\cal E}',{\cal E}} - p I_{O,{\cal E}}  \right)^+ $  \\
\hline \hline
$(n, \ov \nu^{\dot \alpha})$ & $(-2, 2)$ & $ \frac{1}{2} \left( I_{{\cal E}',{\cal E}} + p  I_{O,{\cal E}}  \right)^-  $  \\
$\nu^{\alpha}$ & $-2$ & $ \frac{1}{2} \left( I_{{\cal E}',{\cal E}} - p  I_{O,{\cal E}}  \right)^- $  \\
\hline
\hline
\end{tabular}
\caption{Charged zero modes at an ${\cal E}- {\cal E}'$ intersection, with p=1,2 for E2-instantons in Type IIA and E3-instantons in Type IIB, respectively. 
\label{antizero} }
\end{table}

\subsection{Superpotential calculus}
\label{subsecsuperpot}

In the previous section we have seen that a rigid $O(1)$ instanton
has the appropriate universal zero mode structure $d^4x \,  d^2 \theta$ to yield a contribution to the
holomorphic F-terms in the effective supergravity action.
In this section
  we will review the explicit computation of 
such corrections to the superpotential. As mentioned in the introduction, in lack of a second quantized version of string theory
we cannot derive the instanton calculus from first
principles, but have to define it from analogy considerations
to field theory.

The starting point is to find the interactions of the instanton zero modes appearing in $S^{(\rm int.)}_{\cal E} ({\cal M})$  in equ. (\ref{saturation}). They can be computed in terms of correlators in the 
boundary conformal field theory describing the interactions
of the D-branes and the instanton. One then integrates out the zero modes by pulling down appropriate interaction terms from the exponent.

Equivalently one can view the computation entirely from the CFT perspective: 
To detect a contribution
like (\ref{superpotex})
to the superpotential one computes an M-point correlator
in an instanton background 
$\langle \Phi_{a_1,b_1}\cdot\ldots\cdot   \Phi_{a_M,b_M}  
\rangle_{\cal E}$.  
For canonically normalized conformal fields this yields the physical correlator.
In terms of the quantities in the effective supergravity
action, it involves a combination of
the superpotential coefficient $Y$, the K\"ahler potential ${\cal K}$ and the 
matter field K\"ahler metrics $K_{a_i,b_i}$,
\begin{eqnarray}
      && \langle \Phi_{a_1,b_1}\cdot\ldots\cdot   \Phi_{a_M,b_M}  
\rangle_{\cal E} =  
\frac{e^\frac{{\cal K}}{2}\, Y_{\Phi_{a_1,b_1},\ldots, \Phi_{a_M,b_M}} }{ \sqrt{ K_{a_1,b_1}\cdot\ldots\cdot K_{a_M,b_M}   }}. 
 \label{Yukawaholokaehler}
\end{eqnarray}
Now focus on the superpotential contribution. As for the quantum
fluctuations around the classical instanton solution, 
only terms proportional to $g_s^0$ are relevant as these
translate into a constant dependence 
on the holomorphic superfields
$T_i$ and $U_i$ in eqs. (\ref{superfieldsa}), (\ref{superfieldsb}) and, respectively, 
in eq. (\ref{superfieldsc}). 
Any other dependence
on these axions and therefore on $g_s$ can be ruled out due to the axionic
shift symmetries. 
In conclusion, the counting of factors of $g_s$ together with the need to insert 
all fermionic zero modes
gives the terms which can appear in the
superpotential. 

Now, each disc diagram carries
an overall normalization factor proportional to $ g_s^{-1}$.
Each annulus or M\"obius diagram comes with an additional factor of $g_s$.
In \cite{bcw06,crw07} it was argued, in analogy with the ADHM construction by D-brane instantons \cite{MB02}, that one should assign 
to each  charged fermionic zero mode $\lambda_a$   
an extra  factor of $\sqrt{g_s}$.
From the counting of $g_s$ it is therefore  clear that  only worldsheets with
the topology of a disc or of an annulus or M\"obius strip
can contribute to the superpotential.
Furthermore, charged fermionic zero modes
can only contribute to the disc amplitudes and in such a way that precisely two of them are inserted, whereas
the 1-loop amplitudes have to be uncharged. We will provide more evidence for this picture momentarily.

\begin{figure}[ht]
\centering
  \includegraphics[width=0.2\textwidth]{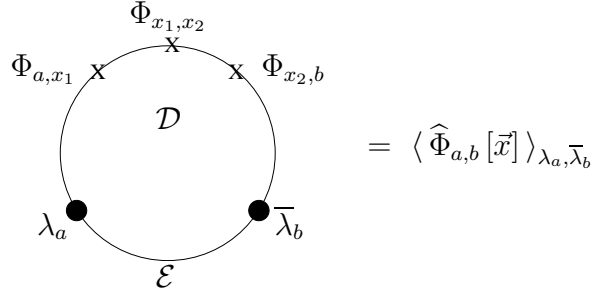}
\begin{picture}(100,1)
\put(-95,10){$\lambda_a$}
\put(-5,10){$\ov\lambda_b$}
\put(-50,-10){${\cal E}$}
\put(-50,50){${\cal D}$}
\put(-105,70){$\Phi_{a,x_1}$}
\put(-60,90){$\Phi_{x_1,x_2}$}
\put(-10,70){$\Phi_{x_2,b}$}
\put(30,40){$=\ \langle\, \widehat\Phi_{a,b}\,[{\vec x}]\,\rangle_{\lambda_a,\ov\lambda_b}$}

\end{picture}
\caption{Disc diagram for charged zero mode absorption.
\label{zeroabsorb}}
\end{figure}

Summarizing all these observation, in \cite{bcw06} the following
formula for computing the correlation function in the semi-classical
approximation\footnote{Higher loop corrections are certainly not vanishing,
but come from corrections to the K\"ahler potential resp. K\"ahler metrics
in (\ref{Yukawaholokaehler}).}
was proposed
\bea
\label{spacetimecorrelator}
      \hskip -1cm \langle \Phi_{a_1,b_1}\cdot\ldots\cdot   \Phi_{a_M,b_M}  
\rangle_{\cal E}  
 &\simeq& \int d^4 x\, d^2\theta \,\, 
       \sum_{\rm conf.}\,\,  {\textstyle  
  \prod_{a} \bigl(\prod_{i=1}^{ I^+_{{\cal E},{\cal D}_a} }  d\lambda_a^i\bigr)\, 
               \bigl( \prod_{i=1}^{ I^-_{{\cal E},{\cal D}_a} }  d\overline{\lambda}_a^i\bigr) } \nonumber \\ 
   && \phantom{a} \exp ({-S^{(0)}_{\cal E}}) \,\, 
           \exp \left(   \sum_b\: \annbb{{\cal E}}{{\cal D}_b}^{\!\star} + 
          \annbc{{\cal E}}{O}^{\!\star}    \right) \,\, \\
&& \langle \widehat\Phi_{a_1,b_1}[\vec x_1]   \rangle_{\lambda_{a_1},\overline{\lambda}_{b_1}}\cdot 
            \ldots \cdot  \langle \widehat\Phi_{a_L,b_L}[\vec x_L] 
          \rangle_{\lambda_{a_L},\overline{\lambda}_{b_L}}\,.\nonumber
\eea
For more details on the overall normalization see \cite{crw07}. The amplitude (\ref{spacetimecorrelator}) involves an integration over all instanton zero modes and a sum over all configurations of distributing the vertex operators
for the charged matter fields $\Phi_{a_i,b_i}$ on disc diagrams, on each of which two charged zero modes are
inserted.\footnote{Note that for simplicity the possibility of including annuli with charged \emph{matter}
fields inserted is neglected.} 
The abbreviation $\widehat\Phi_{a_k,b_k}[{\vec x_k}]$ denotes
a chain-product of vertex operators
\begin{eqnarray}
  \widehat\Phi_{a_k,b_k}[{\vec x_k}] = \Phi_{a_k,x_{k,1}}\cdot  
  \Phi_{x_{k,1},x_{k,2}} \cdot \Phi_{x_{k,2},x_{k,3}}\cdot \ldots 
  \cdot \Phi_{x_{k,n-1},x_{k,n}}  \cdot \Phi_{x_{k,n},b_k} \quad ,
\end{eqnarray}
while 
$\langle \widehat\Phi_{a_1,b_1}[\vec x_1]   \rangle_{\lambda_{a_1},\overline{\lambda}_{b_1}}$ is a CFT disc correlator
with the vertex operators for $\widehat\Phi_{a_1,b_1}[\vec x_1]$ and those for the charged zero modes
$\lambda_{a_1}$ and $\overline{\lambda}_{b_1}$ inserted on the boundary. 
Therefore, the charged matter zero modes are soaked up
by boundary changing CFT disc correlators as shown in figure \ref{zeroabsorb}. They were computed explicitly in \cite{crw07}.
\vspace{0.3cm}

Note that the respective arguments of the two exponential functions in
(\ref{spacetimecorrelator}) are the disc vacuum diagram
and the one-loop vacuum diagram with at least one
boundary on the instanton ${\cal E}$. 
The vacuum disc amplitude for an Ep-instanton is given by \cite{JP94}
\bea
\label{Szero}
   S^{(0)}_{{\cal E}}= -\langle 1 \rangle^{\rm disc}
    ={1\over g_s}{V_{\cal E}\over 
    \ell_s^{p+1}}={8\pi^2\over g_{\rm YM, {\cal E}}^2}\,.
\eea
Here $g_{\rm YM, {\cal E}}$ is the gauge coupling on an auxiliary spacetime-filling D-brane ${\cal D}_{\cal E}$ that would be wrapping the same internal cycles as the instanton ${\cal E}$. In particular, this quantity is not identical to the gauge couplings on the branes ${\cal D}_{a,b}$. 

\begin{figure}[ht]
\centering
\hspace{40pt}
\includegraphics[width=0.6\textwidth]{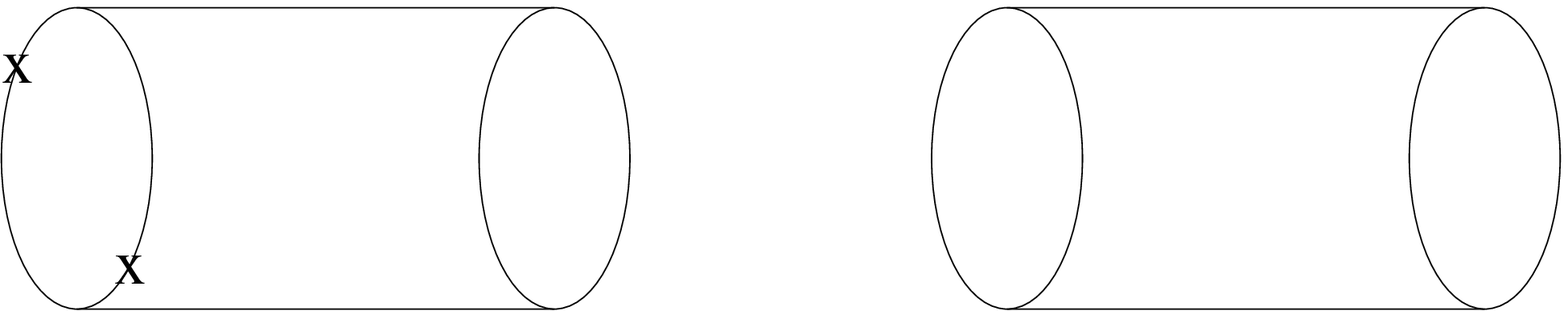}

\begin{picture}(100,1)
\put(-47,38){${\cal D}_{\cal E}$}
\put(29,38){${\cal D}_b$}
\put(62,38){$=\, {\rm \Re}\Biggl[$}
\put(199,38){$\Biggr]$}
\put(103,38){${\cal E}$}
\put(176,38){${\cal D}_b$}
\put(-62,50){$F$}
\put(-26,20){$F$}

\end{picture}

\vspace{-10pt}
\caption{Annulus 1-loop vacuum diagram.\label{figannu}}
\end{figure}

For the one-loop diagrams $\annsbb{{\cal E}}{{\cal E}}=0$
and $\annsbb{{\cal D}_a}{{\cal D}_b}=0$ by supersymmetry. Thus
only the annulus and M\"obius strip amplitudes with
precisely one boundary on ${\cal E}$ really contribute.
(The upper index $^\star$ in (\ref{spacetimecorrelator}) 
will be explained momentarily.)
In \cite{ag06,ablps06} it was shown that these vacuum diagrams
are related to the gauge threshold corrections associated with the  auxiliary D-brane
 ${\cal D}_{\cal E}$.
In stringy Feynman diagrams we therefore arrive at the
relation shown in figure \ref{figannu}.

\noindent
Analogously, one finds a relation for the instanton vacuum
M\"obius strip amplitudes
$\anntc{{\cal D}_{\cal E}}{O}=\Re[\annbc{{\cal E}}{O} ]$.
\vspace{0.1cm}

All disc and 1-loop CFT amplitudes  
in the instanton correlator (\ref{spacetimecorrelator})
factorize into holomorphic and non-holomorphic parts.
For disc amplitudes a formula completely analogous
to (\ref{Yukawaholokaehler}) holds and for the annulus and M\"obius
strip amplitudes  one employs the Kaplunovsky-Louis formula 
\bea
\sum_b\: \Re\bigl[ \annbb{{\cal E}}{{\cal D}_b} \bigr] +
         \Re\bigl[ \annbc{{\cal E}}{O}\bigl] &=&   
-8\pi^2\,\mathrm{Re} ( {f^{(1)}_{\cal E}}) - \frac{\beta}{2}\log\left( \frac{M_{p}^2}{ 
    \mu^2}\right) - \frac{\gamma}{2}\, {\cal K}_{\rm tree}\\
  &&-\log\left( \frac{V_{\cal E}}{g_s}\right)_{\rm tree} + \sum_b
  \frac{|I_{{\cal E}, {\cal D}_b} N_b|}{2} 
\log\left[\det K^{{\cal E}{\cal D}_b} \right]_{\rm tree}\, . \nonumber
\label{KLinstanton}
\eea
Here ${f^{(1)}_{\cal E}}$ denotes the holomorphic piece 
from the annulus diagram (Wilsonian one-loop threshold correction).
For the brane and instanton configuration in question the coefficients 
are   
\begin{eqnarray}
   \beta=\sum_b \frac{|I_{{\cal E},{\cal D}_b} N_b|}{2} -3,\quad\quad  
   \gamma=\sum_b \frac{|I_{{\cal E},{\cal D}_b} N_b|}{2} -1. 
\end{eqnarray}
The sum over the ${\cal D}_b$ branes include also the
$\Omega_p$ image branes.
The coefficient $\beta$ is nothing else than 
the one-loop $\beta$-function coefficient for
the gauge theory of an auxiliary  ${\cal D}$ brane wrapping
the same cycle as ${\cal E}$.
It involves the one-loop correction of massless modes.
However, since in (\ref{spacetimecorrelator}) the integral over these zero modes
is carried out explicitly, to avoid double counting
we have to remove their contribution from (\ref{KLinstanton}).
This is the definition of the $^\star$  in (\ref{spacetimecorrelator}).
Consistently, the zero mode measure leads precisely to a divergence
\bea
      \mu^{{N_f\over 2}-N_b}=\mu^{\beta}
\eea
with $N_f$ denoting the total number of fermionic zero modes
and $N_b$ the number of bosonic zero modes.

It was shown in \cite{Akerblom:2007uc} that a number of  cancellations appear,
which indeed allow one to express the holomorphic superpotential coupling
entirely in terms of the holomorphic couplings in the
CFT amplitudes 
\bea
     Y_{\Phi_{a_1,b_1},\ldots, \Phi_{a_M,b_M}}&=& 
       \sum_{\rm conf.}\, \, \exp \left(-S^{(0)}_{\cal E}\right) 
           \,\,  \exp \left(-f^{(1)}_{\cal E}\right)\, \\ 
       &&\quad\quad  Y_{\lambda_{a_1}\, \widehat\Phi_{a_1,b_1}[\vec x_1]\, \overline\lambda_{b_1}} 
\cdot\ldots\cdot Y_{\lambda_{a_1}\, \widehat\Phi_{a_L,b_L}[\vec x_L]\, 
         \overline\lambda_{b_L}} .\nonumber
 \label{Yukawaholo2}
\eea
This can be considered a non-trivial consistency check
of the instanton calculus. In particular the one-loop vacuum
amplitudes and the rule that only two charged
zero modes are attached to each disc play a crucial role.
The latter prescription is also a consequence of the following observation: If one replaces the instanton by a spacetime-filling D-brane the vertex operators of the charged zero modes will correspond to fermionic fields, and disc amplitudes with more than two fermions do not give rise to holomorphic contact terms.  
While these arguments make it clear that discs with more than two charged zero mode insertions do not yield superpotential terms, it would be interesting to further investigate their role.

In the sequel we will often denote by
\bea
S_{\cal E} = S^{(0)}_{\cal E} + f^{(1)}_{\cal E}
\eea
the tree-level plus one-loop holomorphic piece of the instanton suppression factor. 

Let us mention that the Kaplunovsky-Louis formula 
(\ref{KLinstanton}) can also be applied to extract information on the
non-holomorphic quantities, i.e. in particular on the matter field
K\"ahler metrics. This was carried out for intersecting
D6-branes in various toroidal orbifolds in 
\cite{abls07b,abls07a,MP07a,MB07b,bs07}.

\section{GENERATION OF F-TERMS: OVERVIEW}
\label{sGenerationFterms}

Based on this instanton calculus, a lot of recent
work has been devoted to its applications and generalizations.
This section aims at providing a guide through
and a logical ordering of the extensive literature.
Since we will elaborate on phenomenological applications to
string model building with Type II orientifolds in
\S\ref{sphenoappl}, here we discuss other interesting, but slightly more formal developments.
A summary of the various effects associated with BPS instantons of different types  is given in table \ref{tab_OverviewF}.

\begin{table}[ht]
\centering
\begin{tabular}{|c|c|c|c|c|}
\hline
instanton     & parallel       & universal  & extra  fermionic      & effective   \\
gauge group   & branes         & zero modes & zero modes            & contribution   \\
\hline \hline
              &                 &           & -            & superpotential    \\ 
$ {\cal O}(1)$&  -             &     $x^{\mu}, \theta^{\alpha}$&  $\gamma^{\alpha}$  & gauge kin. function \\ 
              &                &              & $\ov \chi^{\dot\alpha}$ & multi-fermion for vectors \\[1ex] \hline
              &  -             &              & -           &  multi-fermion  for hypers \\
              &                &              &            & multi-fermion for vectors/ \\ [-1.5ex]
  $U(1)$        &  $\raisebox{1.5ex}{-}$        &    $x^{\mu}, \theta^{\alpha}, \ov \tau^{\dot \alpha}$          &     $\raisebox{1.5ex}{ $\ov \mu^{\dot \alpha}, \ov \nu^{\dot \alpha}$  }$      
   & superpotential \\
              &1                 &             & ADHM          &  superpotential \\
              &  $ N_c > 1$      &            & ADHM         &  gauge instanton \\
\hline
\end{tabular}
\caption{ Overview of F-term generation by BPS instantons in absence of background flux.}
\label{tab_OverviewF}  
\end{table}

\subsection{Gauge instanton effects}
\label{sgaugeinst}

\subsubsection*{ADS superpotential for $SU(N_c)$ SQCD}

Historically the string theoretic derivation \cite{EW95b,MD95} of the famous ADHM construction of gauge instantons \cite{ahdm78} was among the first appearances of Euclidean D-branes in the context of instanton computations and has been the subject of many early investigations including \cite{gg97a,gg97b,MB02} (for reviews and more references see \cite{bvv00,dhkm02,bkr07}).

Here we will focus on well-known non-perturbative effects in ${\cal N}=1$
supersymmetric gauge theories \cite{is95}. It is interesting
to verify that indeed the D-brane instanton calculus
contains these effects. 
In this section, we provide some of the details of how
this works for the prototype example of $SU(N_c)$ ${\cal N}=1$ supersymmetric QCD
with $N_f=N_c-1$ flavors. In this supersymmetric field theory
a gauge instanton generates the so-called Affleck-Dine-Seiberg
(ADS) superpotential \cite{ads83}
\bea
 S_W=\int d^4 x\, d^2\theta \,\,   
        {\Lambda^{3N_c-N_f} \over  {\det}[M_{ff'}] },
\eea
where $M_{ff'}$ is the meson matrix and $\Lambda$ the dynamically
generated scale.
The way to proceed is to engineer a local D-brane set-up
realizing this situation and to describe the gauge instanton as
a D-brane instanton. One then computes the instanton
amplitude following the rules explained in \S\ref{subsecsuperpot}
and takes the field theory limit.

\begin{figure}[ht]
\centering
\hspace{40pt}
\includegraphics[width=0.25\textwidth]{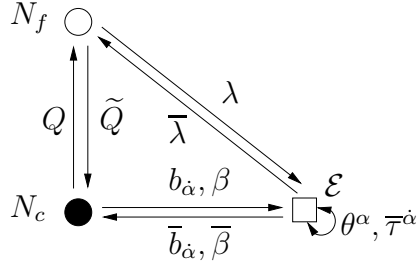}

\begin{picture}(100,1)
\put(119,30){${\cal E}$}
\put(0,22){${N_c}$}
\put(0,95){${N_f}$}
\put(12,55){$Q$}
\put(34,55){$\widetilde{Q}$}
\put(80,65){$\lambda$}
\put(60,50){$\ov\lambda$}
\put(60,32){${b_{\dot \alpha},\beta}$}
\put(60,8){${\ov{b}_{{\dot\alpha}},\ov\beta}$}
\put(125,15){$\theta^\alpha, \ov\tau^{\dot\alpha}$}

\end{picture}

\vspace{-10pt}
\caption{Extended quiver diagram showing the matter fields 
  and the instanton zero modes for $SU(N_c)$ SQCD with $N_f$ flavors. 
\label{figa}}
\end{figure}

To engineer ${\cal N}=1$ SQCD we wrap a stack of $N_c$ D-branes on a rigid
cycle and $N_f$ D-branes on another cycle such that  the
intersection realizes precisely one vector-like pair of matter fields
$Q$ and $\widetilde{Q}$
transforming in the bifundamental representation $(N_c,N_f)$.\footnote{Such configurations can also be engineered in quiver 
gauge theories arising from fractional D-branes on singularities. More on that in section \ref{secquiver}.}
The gauge instanton (in the zero size limit) is described in string
theory by a Euclidean D-brane ${\cal E}$ wrapping the
same internal cycle as the color brane ${\cal D}_c$. 
The final local ${\cal D}-{\cal E}$  brane configuration and the 
resulting zero modes are shown  in figure \ref{figa}.

\noindent
Let us discuss the appearing zero modes in some more detail:
\vspace{0.3cm}

\noindent
\underbar{${\cal E}$-${\cal E}$}: 
Since the ${\cal E}$-instanton wraps the
same rigid cycle as the ${\cal D}_c$ brane it is
a $U(1)$ instanton. Its universal 
${\cal E}$-${\cal E}$ zero modes are 
the four positions in Minkowski space $x^\mu$ with $\mu=0,\ldots,3$ and 
the four fermionic zero modes  $\theta^\alpha$, $\ov\tau^{\dot\alpha}$.

\vspace{0.3cm}
\noindent
\underbar{${\cal E}$-${\cal D}_f$}: In this sector, one only gets the 
$N_f$ pairs of non-chiral $\lambda_f, \ov\lambda_f$ zero modes
from Table \ref{tablezero}.

\vspace{0.3cm}
\noindent
\underbar{${\cal E}$-${\cal D}_c$}: This sector, which has not been discussed
in \S\ref{subsec_zero}, is characteristic of gauge instantons. 
Since the ${\cal E}$-instanton and the ${\cal D}_c$ branes wrap 
the same cycle, open strings between them are subject to
six $NN,DD$ boundary conditions and four $DN$ boundary conditions. 
Therefore, the ground state energy in both the NS- and the R-sector 
vanishes and one finds $4 N_c$ bosonic zero modes $b^u_{\dot \alpha}$, 
$\ov b_{\dot\alpha,u}$
and $2 N_c$ fermionic ones $\beta^u, \ov\beta_u$ with $u=1,\ldots, N_c$.

However, not all of these  zero modes are independent. 
In fact, as shown in \cite{MB02} the effective action on the
${\cal E}$ contains the two terms
\bea
\label{ads1}
     S_{1}=i\, \ov\tau^{\dot\alpha} \left( b^u_{\dot\alpha}\, \ov\beta_u
                + \ov{b}_{\dot\alpha,u}\, \beta^u \right)-i\,
                    D^c \left( \ov{b}^{\dot\alpha}_u\,\,
                      (\tau_c)^{\dot\beta}_{\dot\alpha}\,\,
              b^u_{\dot\beta} \right).
\eea
Here $D_c$ with  $c=1,2,3$  are  auxiliary fields which, together with   
the extra Goldstinos $\ov \tau^{\dot \alpha}$, appear as Lagrangian multipliers implementing
the D- and F-term constraints in the effective action on the instanton.
Integrating out  $\ov\tau^{\dot\alpha}$ and $D^c$ yields
precisely the fermionic and, respectively, bosonic ADHM constraints for the
case of a single instanton,
\bea
 b_{\dot\alpha}\, \ov\beta
                + \ov{b}_{\dot\alpha}\, \beta = 0, \quad \quad\quad \ov{b}^{\dot\alpha}\,\,
                      (\tau_c)^{\dot\beta}_{\dot\alpha}\,\,
              b_{\dot\beta} = 0.
\eea
 Note that  the universal
zero modes $\ov\tau^{\dot\alpha}$ are soaked up in this process.
This is the microscopic reason why an ${\cal E}$ brane
instanton on top of a ${\cal D}$ brane can contribute to the
holomorphic superpotential
\bea
           S=\int d^4 x\, d^2\theta\  W
\eea
with 
\bea
\label{adssuppe}
  W={\cal C}\, \int d\{ b^u_{\dot\alpha}, \ov{b}_{\dot\alpha,u},
                 \beta^u, \ov\beta_u, \lambda_f,\ov\lambda_f\}\,\,\,
          \delta(  b_{\dot\alpha}\, \ov\beta
                + \ov{b}_{\dot\alpha}\, \beta )\,\,
          \delta(\ov{b}^{\dot\alpha}\,\,
                      (\tau_c)^{\dot\beta}_{\dot\alpha}\,\,
              b_{\dot\beta} )\, \,\,
              e^{-S_{\cal E}-S_{2}}.
\eea
Here in the two fermionic and three bosonic ADHM constraints summation
over the color index is understood. 

One now has  to soak up the remaining fermionic zero modes in such a way that 
the bosonic integral converges. The computation of the
appropriate three-point and four-point disc amplitudes
yields couplings \cite{ablps06}
\bea
     {S}_{2} =  \beta^u\,\, (Q^\dagger)_u^f \,\, \ov\lambda_f
           + \lambda^f\,\, (\widetilde{Q}^\dagger)^u_f\,\, \ov\beta_u
 +  {1\over 2}\,  \ov{b}_{\dot\alpha,u} \left( Q^u_f\,\, 
              (Q^\dagger)^f_v +(\widetilde{Q}^\dagger)^u_f\,\,
                 \widetilde{Q}^f_v  \right)
               b^{\dot\alpha,v}.
\eea
Note that it is really the anti-holomorphic fields $Q^\dagger$ and
$\widetilde{Q}^\dagger$ which enter into these couplings.
Inserting this action into eq. (\ref{adssuppe}), one can now
compute the resulting integrals. 
In view of the two fermionic ADHM constraints, a simple counting argument yields that only for $N_f=N_c-1$
the fermionic zero mode integral is non-vanishing.\footnote{For  $N_f \geq N_c$ the remaining zero modes have to be absorbed by additional interactions which will be discussed in \S\ref{sec_BW} and lead to higher fermion F-terms instead of a superpotential.}
After first integrating over the fermionic zero modes, one
is left with a Gaussian integration over the bosonic ones.
These integrals can be carried out as detailed for instance
in \cite{ablps06}. The D-terms for the $SU(N_c)$ gauge theory
constrain the vevs of the quark fields such that
$Q\,Q^\dagger=\widetilde{Q}^\dagger\, \widetilde{Q}$. This indeed leads
 to a cancellation of the anti-holomorphic terms.
One eventually arrives at the  ADS superpotential
\bea
             W=   { M_s^{2N_c+1}\over {\rm det}(\widetilde{Q}\, Q) }
             \,\,  \exp \left( {-{8\pi^2\over g^2_{c}(M_s)}}\right)=
           {\Lambda^{3N_c-N_f} \over  {\det}[M_{ff'}] },
\eea 
where we have introduced the correct dimensionfull scale
and, in the field theory limit, have  neglected all
contributions from massive modes in the 
vacuum one-loop diagrams. 
The dynamically generated scale $\Lambda$ is defined as
\bea
\label{dynlam}
      \left( {\Lambda\over \mu}\right)^{3N_c-N_f} = \exp \left( {-{8\pi^2\over
          g^2_{c}(\mu)}}\right).
\eea 

Very similar computations can be performed for ${\cal N}=1$
SQCD like theories with gauge groups $SO(N_c)$ and 
$SP(2N_c)$ with $N_f=N_c-3$ and $N_f=N_c$ flavors, respectively.
Here the engineering of the gauge theory requires the
introduction of orientifold planes \cite{ablps06}. 
Various generalizations of such local ${\cal N}=1$ quiver type gauge 
theories and the respective modelling of gauge instanton effects by Euclidean D-branes have been discussed in the literature \cite{bk07,abflp07,bfm07,afp08,MB08b,fp09}. A prototype of such geometries will be explained in \S\ref{secquiver}.  \\

\subsubsection*{Stringy instantons for the special case $N_c=1$ and generalizations}

The ADHM computation performed around equ. (\ref{adssuppe}) admits  an interesting application  beyond proper gauge instantons. In fact, the absorption of the $\ov \tau^{\dot \alpha}$ modes with the help of the bosonic and fermionic zero modes in the ${\cal D}_c - {\cal E}$ sector works even in the special case $N_c=1$ \cite{CP07}. This describes an instanton along  a cycle not-invariant under the orientifold action that is wrapped by a single spacetime-filling D-brane.
From the point of view of the abelian gauge theory along this cycle the instanton effect should not be interpreted as a gauge instanton since a $U(1)$ theory does not lead to any strong gauge dynamics. 
Still, after absorption of the $\ov \tau^{\dot \alpha}$ modes and performing the bosonic moduli integral the instanton can generate a superpotential term \cite{CP07,fp09} - provided a contribution is not prohibited by additional modes such as those arising in the ${\cal E} - {\cal E'}$ sector. This effect had been anticipated by quite different methods in \cite{abk07} and will be analysed more closely in \S\ref{sec_GT} and \S\ref{sec_RG}.

In \cite{hmssv08, mss08b} it was proposed that this reasoning holds true even in a more general situation. These papers consider a single E3-instanton wrapping the same cycle as a D7-brane, but carrying in addition non-trivial gauge flux ${\cal F}_{E3} \neq {\cal F}_{D7}$. In this case the bosonic and fermionic modes $b_{\dot \alpha}, \beta$ in the E3-D7 sector are massive due to the twisting by the relative gauge flux. However, it was argued that couplings of the form (\ref{ads1}) involving their Kaluza-Klein partners, i.e. $\ov \tau^{\dot \alpha} \, (b^{KK}_{\dot \alpha} \, \ov \beta^{KK} + \ov b^{KK}_{\dot \alpha} \, \beta^{KK}  )$, can still be used to saturate the $\ov \tau$-modes as to generate a superpotential. Note that this is in deviation from our previous policy to include the massive modes, which are off-shell, only in the one-loop factors and not in the instanton effective action. It will be interesting to further verify if such configurations, which were also used in the phenomenological applications of \cite{mss08a,hv08b}, really contribute to the superpotential.

\subsection{Corrections to the gauge kinetic function}
\label{sec_gaugekin}

Up to now we have discussed instanton induced corrections to the
superpotential. In ${\cal N}=1$ supergravity there
exists another holomorphic quantity, namely the gauge kinetic function
appearing in
\bea
               S_{\rm Gauge}=\int d^4 x\, d^2 \theta \,\,
                   f(T,U)\,\, {\rm tr}\, W^\alpha\, W_{\alpha}.
\eea
Here we consider the gauge fields to come from space-time
filling intersecting ${\cal D}_a$ branes.
In \S\ref{subsecDinst} we have already discussed that $f(T,U)$
only receives perturbative corrections up to one-loop order 
beyond which only  D-brane instanton corrections are allowed.
Moreover, we discussed the possible dependence on the complex
structure $U$ and K\"ahler moduli $T$.

Again $O(1)$ instantons have the correct universal zero mode 
structure to yield a non-vanishing contribution
to this F-term.  But the two zero modes $\theta^{\alpha}$ alone are not sufficient  to generate a non-vanishing instanton amplitude.
It turns out that one needs precisely one pair
of deformation zero modes of the type $\gamma^\alpha$ \cite{abls07b} as listed in the first line of table \ref{tabledeform}.
In addition there must be no charged zero modes from intersections with other D-branes.
For a discussion of analogous corrections in heterotic string theory by worldsheet instantons wrapping
higher genus curves see \cite{bw05}.

\begin{figure}[ht]
\centering
\hspace{10pt}
\includegraphics[width=0.3\textwidth]{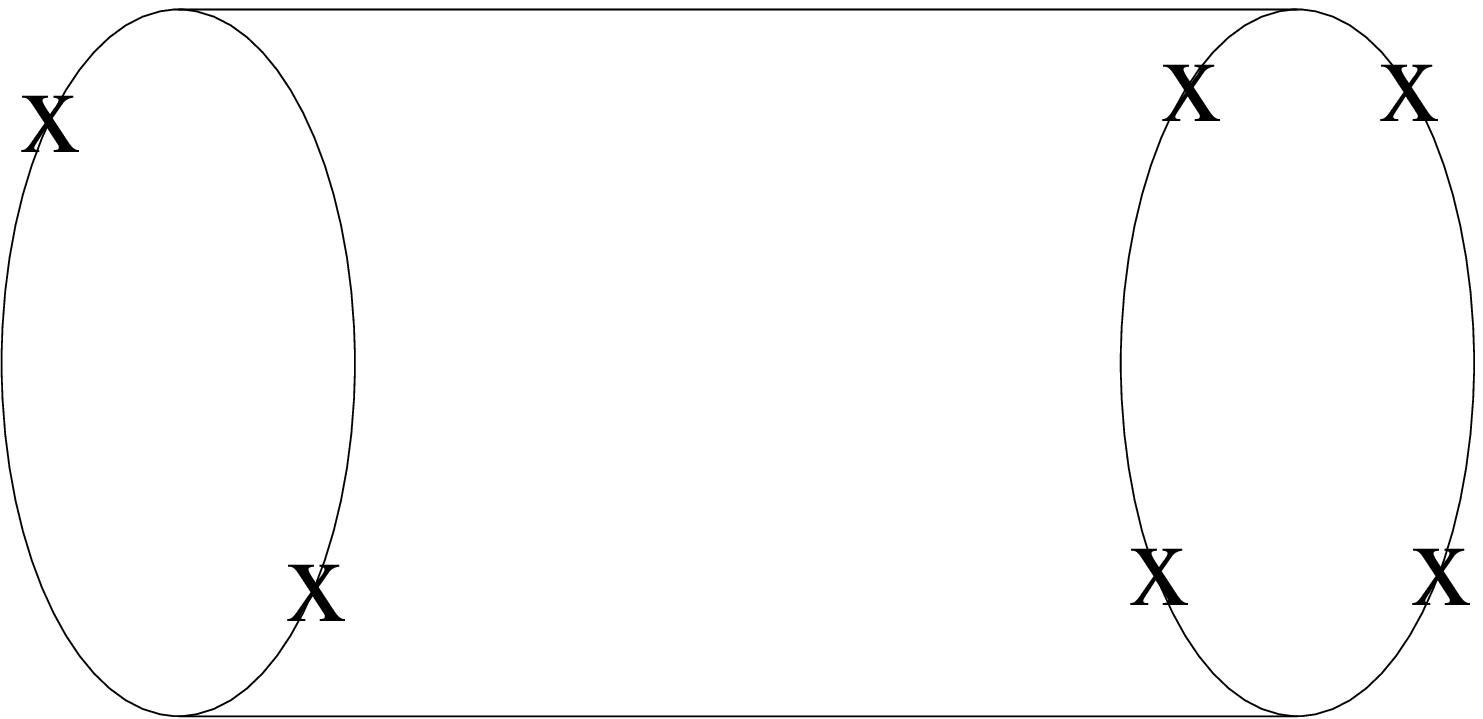}

\begin{picture}(100,1)
\put(4,42){${\cal D}_a$}
\put(100,42){${\cal E}$}
\put(-16,63){$F_a$}
\put(23,22){$F_a$}
\put(78,61){$\theta^{1\over 2}$}
\put(76,22){$\theta^{1\over 2}$}
\put(119,64){$\gamma^{-{1\over 2}}$}
\put(119,24){$\gamma^{-{1\over 2}}$}
\end{picture}

\vspace{-10pt}
\caption{Annulus diagram for an ${\cal E}$-instanton correction to 
the gauge coupling of a stack of ${\cal D}_a$ branes. 
The upper indices give the ghost number of the vertex operators.\label{finst}}
\end{figure}

The  four fermionic zero modes $\theta^{\alpha}, \gamma^{\alpha}$ of the instanton 
can be absorbed by an annulus diagram as shown in figure \ref{finst}.

\noindent
The total amplitude that computes an ${\cal E}$-instanton
correction to the gauge kinetic function $f_a$ has
the form
\bea 
\label{gaugeka} 
    f^{\cal E}_a=\int d^2\theta \, d^2\gamma\  \,\, \anntf{{\cal D}_a}{\cal E}\
    \, e^{  - S^{(0)}_{\cal E}}
    \,\, \exp\!\left(  \sum_b\: \annbb{\cal E}{{\cal D}_b} + 
        \annbc{\cal E}{O}
    \right) \; .
\eea 
The last factor represents  the exponentiated disc and
one-loop vacuum diagrams with at least one boundary on the
${\cal E}$ instanton, as is by now familiar from the superpotential calculus.

Next we  need to know  the zero mode
absorption amplitude between the ${\cal D}_a$-brane and
an ${\cal E}$-instanton. It was shown in \cite{bs08} 
that this at first sight highly complicated six-point function
can be related to the derivative of a much easier 
two-point function. In fact it is the derivation of a 
one-loop gauge threshold correction with respect to
the deformation moduli $m$ 
\bea
\label{strangeg}
   \Re \left[ \int d^2\theta d^2\gamma \  \anntf{{\cal D}}{\cal E} \right]
      ={\partial^2\over \partial {m}^2}
 \ \annbt{{\cal D}}{{\cal D}_{\cal E}}\biggr\vert_{m=m_0}, \; 
\eea
where again ${\cal D}_{\cal E}$ denotes an auxiliary 
space-time filling ${\cal D}$ brane wrapping the same internal
cycle as the instanton brane ${\cal E}$.

Such single instanton corrections to the gauge kinetic function
were evaluated for a simple toroidal orientifold model in \cite{bs08}.
These results were compared to worldsheet instanton corrections
in an S-dual heterotic string model \cite{cdmp07} (see also \cite{cd08})and complete agreement
for the single instanton contributions were found.
It was pointed out in \cite{bs08} that the existence of instanton
corrections to the gauge kinetic function of D-branes leads
to an iterative structure. This can be interpreted
as multi-D-brane instanton corrections to $f_a$, where
the additional  zero modes are absorbed among the instantons
themselves. More on that will be presented in \S\ref{subsecmultiinst}
on multi-instanton corrections. For a different recent test of \emph{six-dimensional} Type I - Heterotic duality with the help of stringy instantons see \cite{bm07}.

\subsection{Beasley-Witten F-terms}
\label{sec_BW}
There exists yet another class of supersymmetric F-terms in addition to the familiar superpotential and gauge kinetic function.  At the fermionic level these interactions involve a product of $2n$ anti-chiral Weyl fermions beyond the chiral fermion bilinear characteristic of a superpotential. They are therefore often called multi-fermion or higher derivative F-terms. Such interactions were first studied systematically by Beasley and Witten, originally in connection with  ${\cal N}=1$ supersymmetric QCD with gauge group $SU(N_c)$ and $N_f \geq N_c$ (in the special case $N_c=2$ in \cite{bw04}) and more generally in the context of heterotic worldsheet instantons in \cite{bw05}.

In superspace notation, multi-fermion F-terms can be written as
\begin{equation}
\label{BW_gen}
S = \int d^4 x \, d^2 \theta \, \, w_{\ov i_1, \ldots, \ov i_n, \, \ov j_1, \ldots  \ov j_n}\, (\Psi) \, \left({\ov{\cal D}}^{\dot \alpha} {\ov \Phi}^{\ov i_1 }  {\ov{\cal D}}_{\dot \alpha} {\ov \Phi}^{\ov j_1 }\right)  \ldots \left( {\ov{\cal D}}^{\dot \alpha} {\ov \Phi}^{\ov i_n }  {\ov{\cal D}}_{\dot \alpha} {\ov \Phi}^{\ov j_n } \right).
\end{equation}
Here the degrees of freedom assembled in the chiral superfield $\Phi = \varphi + \theta^{\alpha} \psi_{\alpha}$ appear in the combination
\bea
\label{covderiv}
 {\ov{\cal D}}^{\dot \alpha} {\ov \Phi} = \ov \psi^{\dot \alpha} + \theta_{\alpha} \, (\sigma^{\mu})^{\dot \alpha \alpha} \,
 \partial_{\mu} \ov \varphi.
\eea
Even though this is not manifest in equ. (\ref{BW_gen}), multi-fermion F-terms are supersymmetric if $\omega$ depends holomorphically on some chiral superfields $\Psi$ and is antisymmetric in the $\ov i$ and $\ov j$ indices separately and symmetric under their exchange. In addition $\omega$ is subject to a certain equivalence relation discussed in detail in \cite{bw05} which ensures that (\ref{BW_gen}) cannot be written globally as a D-term. 
More background on the geometric interpretation of higher F-terms can be found in \cite{bw04,bw05}.
For example the case $n=1$, corresponding to a four-fermi interaction, is of the form that describes quantum deformations of the moduli space of ${\cal N}=1$ supersymmetric QCD \cite{NS94} with $N_f=N_c$ \cite{bw04}. This concept of encoding deformations of the moduli space of ${\cal N}=1$ supersymmetric theories in higher F-terms is more general \cite{bw05}. 

At a technical level, multi-fermion F-terms of degree $n$ are generated by a BPS-instanton whose zero modes comprise $n$ extra anti-chiral Weyl spinors, denoted collectively as $\ov \mu_{\ov i}^{\dot \alpha}$, $\ov i = 1, \ldots,n$, which couple in the instanton effective action as
\bea
\label{BW_1}
S_{B.W.} = (\ov \mu_{\ov i})^{\dot \alpha} \, {\ov{\cal D}}_{\dot \alpha} {\ov \Phi}^{\ov i}.
\eea 
Integrating out these extra anti-chiral zero modes pulls down corresponding powers of  ${\ov{\cal D}}_{\dot \alpha} {\ov \Phi}$. Depending on the nature of the spinors $\ov \mu_{\ov i}^{\dot \alpha}$ one obtains
\begin{itemize}
\item deformations of the moduli space of SQCD with gauge group $SU(N_c)$ and $N_f \geq N_c$ (and generalizations thereof) from gauge instantons; here the $(\ov \mu_{\ov i})^{\dot \alpha}$ are the bosonic modes $b^{\dot \alpha}$, $\ov b^{\dot \alpha}$ in the ${\cal E}-D_c$ sector, 
\item deformations of the vector multiplet moduli space from stringy ${\cal O}(1)$ instantons with  $n$ deformation modes $(c,\ov \chi^{\dot \alpha})$, or
\item deformations of the hypermultiplet moduli space from stringy isolated $U(1)$ instantons due to the universal zero modes $\ov \tau^{\dot \alpha}$.
\end{itemize}
We now discuss the different cases in turn.

\subsubsection*{Gauge instantons}

Consider again the microscopic realization of a gauge instanton in ${\cal N}=1$ SQCD with gauge group $SU(N_c)$ and $N_f$ flavors as introduced in \S\ref{sgaugeinst}. For $N_f = N_c -1$ such an instanton reproduces the ADS superpotential (\ref{Yukawaholo2}) upon absorbing the fermionic zero modes via the terms (\ref{ads1}) and (\ref{adssuppe}) in the instanton effective action.
As argued in \cite{mpry08} the instanton effective action contains in addition the couplings 
\bea
\label{ads3}
S_3 = \ov b_{\dot\alpha,u} \, (\ov{\cal D}^{\dot\alpha} \widetilde{\ov Q})^u_f \, \lambda ^f + b_{\dot\alpha,u} \,  (\ov{\cal D}^{\dot\alpha} {\ov Q})^u_f \ov\lambda^f.
\eea
The existence of these terms was verified in \cite{MB08b} via an explicit CFT computation for a local $D3-E(-1)$ system on ${\mathbb C}^3/{\mathbb Z}_2 \times {\mathbb Z}_2$.

The point is now that while $N_c-1$ of the $N_f$ pairs of zero modes $\lambda^f$ and $\ov \lambda^f$ are already saturated by the interaction terms
(\ref{ads1}) and (\ref{adssuppe}), for $N_f \geq N_c$ the remaining
$n= N_f - (N_c-1)$ pairs have to be absorbed via the  couplings (\ref{ads3}). 
This pulls down a term $(\ov {\cal D}^{\dot\alpha} \ov Q \, \ov {\cal D}_{\dot\alpha} \widetilde{\ov Q})^n$.

Integrating out the ADHM moduli is more complicated than for $N_f = N_c-1$ due to the appearance of extra factors of bosonic modes $b^{\dot \alpha}$, $\ov b^{\dot \alpha}$ and we refrain from performing this computation here.
E.g. for simplest case $N_f=N_c=2$ the final result is proportional to \cite{mpry08,MB08b}
\bea
S= {\rm tr} (\ov M M)^{-3/2} \, \epsilon^{ijkl} \, \ov {\cal D}_{\dot\alpha} \ov M_{ij} \, {\cal D}^{\dot\alpha} \ov M_{kl} 
\eea
in terms of the meson field $\ov M_{ij} = \epsilon_{uv}\, Q_i^u Q_j^v$, in agreement with the field theoretic derivation of \cite{bw04}. Generalizations to $N_f > N_c$ and to SQCD with gauge group $SP(2N_c)$ and $N_f \geq N_c+1$ can be found in \cite{mpry08,MB08b}.

\subsubsection*{Deformations of the vector multiplet moduli space and open string terms}

Let us move on to the generation of higher F-terms by stringy instantons with additional zero modes other than $d^4 x \, d^2 \theta$.
The first example  of such modes are the subclass of deformation moduli $(c, \ov \chi^{\dot \alpha})$ displayed in the second line of table \ref{tabledeform}.  For heterotic worldsheet instantons analyzed in \cite{bw05} these moduli correspond to instantons along Riemann spheres moving in a family. 

In general the anti-chiral deformation moduli couple to the closed string moduli sitting in those ${\cal N}=1$ chiral multiplets which descend from vector multiplets of the underlying ${\cal N}=2$ compactification \cite{bcrw07}. The respective closed string fields in Type I/Type IIB orientifolds and Type IIA orientifolds are the complex structure and K\"ahler moduli.

To be explicit consider an E2-instanton in Type IIA orientifolds of ${\cal O}(1)$ type with $b_1(E2)=b_1(E_2)_+=1$ and corresponding deformation moduli $(c, \ov \chi^{\dot \alpha})$.
If we schematically denote by ${\cal T} = T + \theta^{\alpha} t_{\alpha} $ the  ${\cal N}=1$ chiral superfields associated with the K\"ahler moduli, then the  $\ov \chi^{\dot \alpha}$-modulini couple in the instanton effective action as 
\bea
\label{BW_4}
S_{B.W.} = \ov \chi^{\dot \alpha}\,  {\ov{\cal D}}_{\dot \alpha} {\ov {\cal T}}.
\eea
This was demonstrated in \cite{bcrw07} by verifiying that 
both the fermionic open-closed disc amplitude $\langle \ov \chi \, \ov t \rangle$ and its superpartner $\langle \theta^{\alpha} \,\, \ov {\chi}^{\dot \alpha} \,\, \ov T \rangle$ are allowed by $U(1)$ worldsheet charge selection rules. Note that the latter  coupling leads to the interaction term $\theta \sigma^{\mu} \ov \chi  \, \partial_{\mu} \ov T$, which indeed combines with $\ov \chi^{\dot \alpha} \ov t_{\dot \alpha}$ into the coupling (\ref{BW_4}).

Integration over the deformation modulus \footnote{The bosonic moduli $c$ decouple from the computation and merely lead to powers of moduli space volume.} yields an F-term of the form (\ref{BW_gen}), 
\bea
\label{BW_2}
S = \int d^4 x \, d^2 \theta \, \, e^{-S_{\cal E}} \,\, f_{\ov i, \ov j} \,  {\ov{\cal D}}^{\dot \alpha} {\ov {\cal T}}^{\ov i} \,  {\ov{\cal D}}_{\dot \alpha} {\ov {\cal T}}^{\ov j}.
\eea
Here $S_{{\cal E}}= S^{(0)}_{{\cal E}} + f^{(1)}_{\cal E}$ denotes again the tree-level plus one-loop corrected action of ${\cal E}$
as in the case of an ordinary superpotential. The information on the concrete vector multiplet moduli appearing in the Beasley-Witten term is encoded in the tensor $f_{\ov i, \ov j}$ (for the case with one deformation modulus) and depends on the geometric details of the setup. For an explicit example on $K_3 \times T_2$  see \cite{AU08}.

Finally, in the presence of suitable charged zero modes the CFT selection rules also allow Beasley-Witten terms involving open string fields in the ${\cal D}_i-{\cal D}_i$ or ${\cal D}_i - {\cal D}_j$ sector of other D-branes \cite{bcrw07}.
This requires instanton couplings of the schematic type
\bea
S_{B.W.} =  \lambda_a \, \ov \chi^{\dot \alpha} \, {\ov {\cal D}}_{\dot \alpha} \, \ov\Phi_{ab} \, \ov \lambda_b 
\eea
and charged zero modes $\lambda_a, \ov \lambda_b$ in the ${\cal E} - {\cal D}_a$ and ${\cal D}_b - {\cal E}$ sector, respectively.

\subsubsection*{Deformations of the hypermultiplet moduli space}

A new phenomenon arises for stringy $U(1)$ instantons along a non-invariant supersymmetric cycle due the presence of the extra anti-chiral Goldstinos $\ov \tau^{\dot \alpha}$. 
 As discussed in \cite{bcrw07}
if these modes are not lifted by any other mechanism, 
such $U(1)$ instantons generate four-fermi interactions of Beasley-Witten type which involve  the chiral fields descending from the hypermultiplets of the underlying ${\cal N}=2$ compactification. 
This is what happens e.g. for an isolated $U(1)$ instanton not intersecting its orientifold image.
The relevant terms lifting the $\ov \tau$-modes in the instanton effective are completely analogous to the ones described for the deformation modes except that they involve the hypermultiplets. For the example of E2-instantons in Type IIA orientifolds with complex structure moduli ${\cal U} = U + \theta^{\alpha} u_{\alpha}$, the instanton effective action schematically contains the coupling \cite{bcrw07}
\bea
\label{BW_6}
S_{B.W.} = \ov \tau^{\dot \alpha} \, {\ov{\cal D}}_{\dot \alpha} {\ov {\cal U}},
\eea
as follows again from general $U(1)$ worldsheet charge selection rules of the open-closed CFT.
In view of the role of the $\theta^{\alpha}$ and $\ov \tau^{\dot \alpha}$ modes as Goldstinos associated with the two different ${\cal N}=1$ subalgebras, cf. table \ref{tabGoldstino}, it was further argued in \cite{gmu08}
that the combination of hypermultiplet moduli appearing in (\ref{BW_6}) is precisely the superfield $\Sigma$ whose bosonic vev controls the Fayet-Iliopoulos term of the instanton cycle and therefore vanishes for an instanton on its BPS locus. Note that $\Sigma$ appears only in the combination $\ov{\cal D} \ov\Sigma$ so that only the derivatives of its bosonic components enter, see equ. (\ref{covderiv}).  
For a rigid E2-instanton the resulting Beasley-Witten term is then of the form
\bea
\label{BW_3}
S = \int d^4 x \, d^2 \theta \, \, e^{-S_{\cal E}} \,\,  \,
{\ov{\cal D}}^{\dot \alpha} {\ov \Sigma }\, \,  {\ov{\cal D}}_{\dot \alpha} {\ov \Sigma}.
\eea

\subsection{Multi-Instanton processes}

Our presentation has so far focused on  single-instanton processes. As anticipated already in \S\ref{subsec_zero} the general situation involves multiple instantons at the same time. 
There are at least two ways in which such multi-instanton corrections are almost forced upon us.
First, in many cases what is described as a single instanton effect in some
regions of moduli space becomes a multi-instanton process for other values of
the closed string moduli \cite{bcrw07,gu07,crw08,gmu08}. 
Second, there also exist configurations where the multi-instanton 
contribution is not
related to the decay of stable BPS instantons into several BPS constituents
\cite{bs08} but arises due to   the iterated effect that stringy instantons
can correct string instanton actions. 

\subsubsection*{Multi-instantons and Instanton recombination}

This is due to the appearance of lines of marginal or threshold stability in closed string moduli space where supersymmetric cycles decay into several cycles. This transforms the single instanton associated with the original cycle into a multi-instanton. The closed string moduli governing this behaviour are the ones descending from the ${\cal N}=2$ hypermultiplets, i.e. the K\"ahler and complex structure moduli for Type IIB and Type IIA orientifolds, respectively.
There also exist configurations where the multi-instanton contribution is not related to the decay of stable BPS instantons into several BPS constituents \cite{bs08}.

\vspace{0,6cm}
\noindent
{\it Instantons across lines of threshold stability}

\vspace{0.4cm}
\noindent
The simplest and possibly most abundant type of configurations of the first kind is that of an $O(1)$ instanton decaying into a  $U(1)$ instanton and its orientifold image. The latter system is really a two-instanton configuration, at least in the upstairs geometry prior to orientifolding. This process was first analysed in \cite{bcrw07}. Other configurations including such beyond instanton-image instanton systems were considered in \cite{gu07}. $O(1)$ instantons that undergo such a decay are sometimes referred to as decomposable, in contrast to isolated $U(1)$ instantons which never merge with their orientifold image into a single invariant instanton.

Following \cite{bcrw07} let us consider this process of instanton recombination of an ${\cal E}-{\cal E}'$ system starting from the locus in hypermultiplet space where the cycle and its image are split.
The recombination moduli are given by the ${\cal E}-{\cal E}'$ modes of table \ref{antizero}. 
To illustrate the point let us consider the situation where we have only one vector-like pair of such zero modes $(m,\ov m, \ov\mu^{\dot\alpha})$ and $(n,\ov n, \ov \nu^{\dot\alpha})$. The universal moduli of the two sectors are identified, and if we assume that the instanton has no further deformation or charged modes we have to cope with the measure
\bea
\int d{\cal M} = \int d^4 x \, d^2 \theta \,d^2 \ov \tau \,\, dm \, d \ov m \, d^2 \ov \mu \, \,  dn \, d \ov n \, d^2 \ov \nu.
\eea
At first sight it seems hopeless to ever generate a superpotential term. In particular the extra Goldstone modes $\ov \tau^{\dot \alpha}$ appear as an obstruction.
The crucial point is, though, that there exist new interaction terms in the instanton effective action that allow us to absorb in particular the extra fermionic modes. 
First to mention is the interaction
\bea
\label{rec-action}
S_{\ov \tau}=  \ov \tau_{\dot \alpha}\left( m \,\ov \mu^{\dot \alpha} -
                n \,\ov \nu^{\dot \alpha} \right),
\eea
whose generic presence was shown in \cite{bcrw07} with elementary CFT methods.
It follows that the $\ov\tau^{\dot\alpha}$-modes absorb one linear
combination of the fermionic zero modes $\ov \mu^{\dot \alpha}$, $\ov \nu^{\dot \alpha}$, bringing down in addition two powers of bosonic modes from the exponents.
The resulting bosonic integral is damped due to the D-term
\bea
\label{S_D1}
S_D = (2 m \ov m - 2 n \ov n - \xi({\cal U}))^2,
\eea 
see also \cite{gu07}. This D-term is in complete analogy with the situation for spacetime-filling $\cal D$ branes. In particular the Fayet-Iliopoulos term $\xi$ depends, as always, on the hypermultiplet moduli and measures the misalignment of the ${\cal E}$ (and ${\cal E}'$) brane with respect to the orientifold. It vanishes for supersymmetric configurations.

In absence of any other interaction terms, we are left with one extra pair of zero modes given by the linear combination of $\mu$ and $\nu$  which do not enter (\ref{rec-action}). In agreement with what we learnt so far, such an ${\cal E}-{\cal E}'$ configuration generates Beasley-Witten terms involving the vector multiplets. There may however be other interaction terms that lift also this remaining combination of modes. In this case the two-instanton does contribute to the superpotential. A particular way to lift these extra modes was proposed in \cite{bcrw07} and involves extra charged zero modes. 
In \cite{gu07} it is argued that there also exist configurations with couplings whose analogue for the D-branes would be derivable from a quartic superpotential term of the form
\bea
\label{W_quart}
W = (M N)^2,
\eea
where $M$ and $N$ denote the chiral superfields corresponding to the  bosonic and fermionic ${\cal E}-{\cal E}'$ modes. Taking into account that only the anti-chiral Weyl spinors $\ov \mu^{\dot \alpha}, \ov \nu^{\dot \alpha}$ survive the orientifold, this leads to bosonic and fermionic terms of the form
\bea
\label{S_quart}
S_{quart} &=& \ov \mu\,  \ov \mu \, \ov n \,\ov n + \ov \nu \, \ov \nu \, \ov m \, \ov m + 2 \, \ov \mu \, \ov \nu \, \ov m \, \ov n, \nonumber \\
S_F &=& |m \,  n \, m |^2 + |n \, m \, m|^2. 
\eea 
In this case, $S_{quart}$ allows also for the absorption of the remaining linear combination of fermionic ${\cal E}-{\cal E}'$ modes, and the configuration generates a superpotential term. 

These considerations fit with the aforemention picture of instanton recombination as follows. 
In absence of the F-term (\ref{S_quart}) for a supersymmetric configuration with vanishing Fayet-Iliopoulos terms the ${\cal E}-{\cal E}'$ system is at threshold with the  bound state formed by condensing the bosonic moduli $m$, $n$ in a D-flat manner.
As is familiar from the context of D-branes the bound state corresponding to $|m| = |n|$ has one deformation modulus if the ${\cal E}$ (and ${\cal E}'$) brane is rigid.  Similarly, if one hypothetically moves in hypermultiplet moduli space the Fayet-Iliopoulos parameter $\xi$ becomes non-zero. Depending on its sign $m$ or $n$ acquires a vev in a D-flat manner. This is the recombination of instantons in different regions of moduli space referred to at the beginning of this subsection. 
Either way the recombined object is an $O(1)$ instanton with one deformation modulus and therefore just of the right kind to generate Beasley-Witten F-terms. By contrast, situations with a quartic superpotential of the form (\ref{W_quart}) describe an ${\cal E}-{\cal E}'$ state at threshold whose bound state, in regions of moduli space with $\xi \neq 0$, is indeed rigid and thus has every right to contribute to the superpotential.
In particular, the computation on the two-instanton locus agrees with the  expectations for the corresponding bound state. As argued in \cite{gu07} this is as it has to be for the superpotential to be a holomorphic function in particular of the hypermultiplet moduli governing the instanton decay/recombination. Indeed, the line of threshold $\xi=0$ decribes a real codimension one surface in hypermultiplet space, and a holomorphic function  cannot jump across such a real surface.
This analysis can be generalised to other multi-instanton configurations at threshold where the individual components are not related to one another by the orientifold action \cite{gu07}. Further concrete examples along these lines appear in \cite{iu07,jp08}.

\vspace{0,6cm}
\noindent
{\it Instantons across lines of marginal stability}

\vspace{0.4cm}
\noindent
In all these examples continuity of the superpotential across lines of threshold stability is guaranteed by the fact that for either sign of the Fayet-Iliopoulos term there does exist a supersymmetric multi-instanton configuration. More generally, however, supersymmetric cycles can actually decay across proper lines of marginal stability with no BPS object of the same charge existing on the other side. The simplest such situation occurs, in the present context, again for an ${\cal E}-{\cal E}'$ system with, however, just a single set of extra zero modes $m, \ov m,  \ov \mu^{\dot \alpha}$. As is apparent from the D-term 
\bea
S_D = (2 m \ov m - \xi({\cal U}))^2,
\eea 
a supersymmetric $O(1)$ instanton exists only in regions of moduli space where $\xi >0$, while for $\xi < 0$ the ${\cal E} - {\cal E}'$ confugration ceases to be supersymmetric. 
Consistently, in this case there exists a microscopic obstruction for the ${\cal E}-{\cal E}'$ system to yield F-terms in the effective action \cite{bcrw07}. The point is that in a globally supersymmetric configuration of this type there necessarily exist extra charged zero modes $\lambda^i$ in the sector between  ${\cal E}$ and some of the  D-branes present in the configuration.
This follows form the net $U(1)_{\cal E}$ charge in the ${\cal E} - {\cal D}$ sector as 
\bea
\label{chargetot}
\sum_i Q_{\cal E}(\lambda^i) &=& \sum_a N_a \, \left( -I_{ {\cal E},{\cal D}_a}^+ + I_{{\cal E}, {\cal D}_a}^-  - I_{ {\cal E}, {\cal D}_{a'}}^+ + I_{{\cal E},{\cal D}_{a'}}^- \right)  \nonumber \\
&=& - \sum_a N_a \,  \left( I_{{\cal E},{\cal D}_a} + I_{ {\cal E},{\cal D}_{a'}} \right) .
\eea
With the help of the tadpole cancellation condition the latter expression can be seen to be proportional to the chiral intersection number $I_{{\cal E}, O6}$ \cite{bcrw07}. According to table \ref{antizero}, for an ${\cal E}-{\cal E}'$ system with modes $m, \ov m, \ov \mu$ this is non-vanishing. Closer inspection reveals that these extra zero modes required for $U(1)_{\cal E}$ invariance of the zero mode measure cannot be lifted perturbatively in the instanton effective action. Their presence thus annhiliates the contribution of the instanton to the superpotential. Again, this microscopic picture fits nicely with the arguments of \cite{gu07} that an instanton undergoing actual decay should not generate holomorphic couplings. Having said this, there do exist more sophisticated multi-instanton setups where the role of the additional instantons is to lift the excess $\lambda^i$ modes \cite{crw08}. In agreement with the general philosophy the line of marginal stability is thereby transformed into a line of threshold stability, and again no instanton relevant for a superpotential can actually disappear from the BPS spectrum. \vspace{5pt} \\
\indent To summarize, we have seen that in favourable circumstances even $U(1)$ instantons in type II orientifolds can contribute to the superpotential despite the appearance of two extra Goldstone modes $\ov \tau^{\dot \alpha}$. For this to be the case it must be possible to interpret the configuration as a two-instanton configuration ${\cal E}-{\cal E}'$ in the geometry before orientifolding. The interactions in the ${\cal E}-{\cal E}'$ sector can then lift the extra Goldstinos. Contributions are possible whenever there exists some region in hypermultiplet moduli space where this two-instanton configuration forms a bound state of $O(1)$ type, and the results in different patches of moduli space agree. By contrast, there do exist isolated $U(1)$ instantons which can never form a bound state with their image. Consistently, such instantons only yield Beasley-Witten type terms of the form (\ref{BW_3}), at least in absence of fluxes or other extra ingredients to lift the $\ov \tau^{\dot \alpha}$ Goldstinos.  \vspace{5pt} \\ 
\indent Our discussion has focused on the microscopic description of (multi-)instantons in Type II orientifolds preserving ${\cal N}=1$ supersymmetry. Eventually one will want to find closed expressions by performing the sum over all multi-instanton contributions. In the context of ${\cal N}=2$ supersymmetric Calabi-Yau compactifications of Type II theory powerful techniques have been developed to capture instanton corrections to the moduli space metric. 
Recent progress in determining instanton corrections to the hypermultiplet moduli space includes \cite{DR06} (for Type IIB) and \cite{sv07,hms07,rstv07,apsv08,SA09} (for Type IIA). Some supergravity techniques even carry over to ${\cal N}=1$ orientifolds, see \cite{TG07} for a recent example.

Finally, the continuity of physical quantities despite jumps in the BPS spectrum across lines of marginal stability  
can be made very concrete in field theoretic settings \cite{gmn08} and put in precise connection with the recent mathematical insights of \cite{ks08}.

\subsubsection*{Power towers of multi-instantons}

Another interesting aspect of the multi-instanton configurations just described was put forward in \cite{gu07}.
The effect of a, say, two-instanton configuration system involving ${\cal E}_1$ and ${\cal E}_2$ can also be interpreted as the single instanton contribution from ${\cal E}_1$ after including non-perturbative corrections $\Delta S^{n.p.}_{{\cal E}_1}$ to the effective action of ${\cal E}_1$ due to  ${\cal E}_2$ (or vice versa).
Already in the simple example of the vectorlike ${\cal E}-{\cal E}'$ system introduced previously we can 
interpret ${\cal E}'$ as generating an effective mass term for the $\ov \tau$-modes of the form
\bea
\label{multi1}
\Delta S^{n.p.}_{\cal E} = \int d^2 \ov \mu \, d^2 \ov \nu \,  d m \, d\ov m \, dn \, d \ov n \, \, {\rm exp}(-S_{\cal E'} - S_{\rm int}(m,n,\ov \mu, \ov \nu, \ov \tau)) = \ov \tau \ov \tau \, e^{-S_{\cal E'}}.
\eea 
Here $S_{{\cal E}'}= S^{(0)}_{{\cal E}'} + f^{(1)}_{\cal E'}$ denotes the tree-level plus one-loop corrected action of ${\cal E}'$ and $S_{\rm int}(m,n,\ov \mu, \ov \nu, \ov \tau))$ is the sum of the interaction terms (\ref{rec-action}), (\ref{S_D1}) and (\ref{S_quart}).  
In this spirit the contribution of the instanton ${\cal E}$ is schematically \cite{gu07}
\bea
\label{multi2}
\int d^4 x \, d^2 \theta \, d^2 \ov \tau \, {\rm exp}(-S_{\cal E} - \Delta S^{n.p.}_{\cal E}) &=&   \int d^4 x \, d^2 \theta \, d^2 \ov \tau \, {\rm exp}(-S_{\cal E} - e^{-S_{{\cal E}'}} \, \ov \tau \ov \tau ) \nonumber  \\
&=& \int d^4 x \, d^2 \theta \, e^{-S_{\cal E} -S_{{\cal E}'}}.
\eea
The final result follows once we take the square root due to the orientifold identification.

This picture was further generalised and extended in \cite{bs08}. The starting point is the observation that the 
exponential suppression factor in an F-term generated by an instanton ${\cal E}$ is given by the Wilsonian gauge kinetic function $f_{\cal E}$ of the hypothetical spacetime-filling D-brane ${\cal D}_{\cal E}$ wrapping the same cycle,
\bea
{\rm exp}(-S_{{\cal E}}) = {\rm exp}{(-f_{\cal E})}.
\eea 
As discussed in \S\ref{sec_gaugekin}, this gauge kinetic function $f_{\cal E}$ is itself corrected, beyond the one-loop thresholds $f_{\cal E}^{(1)}$, by suitable D-brane instantons,
\bea
f_{\cal E} = f_{\cal E}^{(0)} + f_{\cal E}^{(1)} + \sum_{r} f_{\cal E}^{{\cal E}_r}.
\eea 
The non-perturbative  correction $f_{\cal E}^{{\cal E}_r}$ due to another instanton ${\cal E}_r$ is given in equ. (\ref{gaugeka}). Recall that ${\cal E}_r$ has to be an $O(1)$ instanton with precisely two Wilson line moduli $\gamma^{\alpha}_r$. While the threshold  $f_{\cal E}^{(1)}$ is  already included in the instanton calculus  outlined in \S\ref{subsecsuperpot}, it is natural to conjecture that also the non-perturbative corrections $\sum_{r} f_{\cal E}^{{\cal E}_r}$ appear in the full answer. This means that the full exponential suppression factor of a non-perturbative holomorphic F-term is given by
\bea
\label{tower1}
{\rm exp}( -S_{{\cal E}}) = 
{\rm exp} \left( 
      - S^{(0)}_{{\cal E}}  - f_{\cal E}^{(1)} - \sum_r \int d^4 x_{r} \, d^2 \theta_{r} \, d^2 \gamma_{r} \,\, 
     { \annbf{{\cal E}}{{\cal E}_r} }
    \, e^{ -S^{(0)}_{{\cal E}_r} - f_{{\cal E}_r}^{(1)}- \ldots
  }  
\right).
\eea
Here $x_{r}$ are the zero modes associated with the relative position of the instanton ${\cal E}$ and ${\cal E}_r$, and 
$\theta^{\alpha}_{r}, \gamma^{\alpha}_r$ denote the universal and Wilson line
fermionic zero modes of ${\cal E}_r$.
Similar to equ. \eqref{strangeg}, in was shown in \cite{bs08}
that  the holomorphic piece in 
the four-zero mode absorption
amplitude $\annsbf{{\cal E}}{{\cal E}_r} $ can be computed by the second 
derivative with respect to the Wilson line moduli of the  
threshold correction of the corresponding hypothetical
space-time filling D-branes
\bea
\label{strangef}
   \Re \left[ \int d^4 x_{r}\, d^2\theta_r\,  d^2\gamma_r \  \annbf{{\cal E}}{{\cal E}_r} \right]
      ={\partial^2\over \partial {m}^2}
 \ \annbt{{\cal D}_{\cal E}}{{\cal D}_{{\cal E}_r}}\biggr\vert_{m=m_0} \; .
\eea

Let us assume for simplicity that $f_{\cal E}$ receives corrections from just one single instanton ${\cal E}_r$ .
In a somewhat reverse spirit as  around equs. (\ref{multi1}) and (\ref{multi2}) we can expand 
the exponent in terms of ${\cal E}_r$  and interpret equ. (\ref{tower1}) as the sum over various multi-instanton configurations involving ${\cal E}$ and $n$ copies of the instanton ${\cal E}_r$, with $n=0,1,...$,
\bea
\int d{\cal M}\  e^{-S_{{\cal E}}} 
   &=& \int d{\cal M} \ e^{-S^{(0)}_{{\cal E}} - f_{\cal E}^{(1)}} +  \\
  &+&   \int d{\cal M}  \, \int d{\cal M}_r  \ \annbf{{\cal E}}{{\cal E}_r}  \, 
  e^{-S^{(0)}_{{\cal E}_r} -   f_{{\cal E}_r}^{(1)}}  \, 
  e^{ -S^{(0)}_{{\cal E}} - f_{\cal E}^{(1)} }    
+ \ldots \nonumber.
\eea
Here we abbreviated the respective zero mode measures by $d {\cal M}$ and
$d {\cal M}_r$. To distinguish these stringy multi-instantons effects from
ordinary field theory multi-instanton corrections, they
were called poly-instantons in \cite{bs08}. In general the 
world-volumes of the participating string instantons wrap different
cycles of the underlying geometry. 

Let us finally mention that the existence of these poly-instanton
corrections is still puzzling. 
In \cite{cdmp07} a heterotic freely acting
$\mathbb Z_2\times \mathbb Z_2$ orbifold with gauge group
$SO(32)$ was considered
and via standard worldsheet techniques the
one-loop (in $g_s$) gauge threshold corrections were computed.
This model has a proposed S-dual Type I description, for which
 the poly-instanton calculus was explicitly applied in \cite{bs08}.
It is expected that worldsheet instanton corrections 
for the heterotic string are mapped to E1-brane
instanton corrections in the Type I string.
Indeed, the heterotic result exhibits the  usual sum over
single worldsheet instantons multiply covering
toroidal curves of the internal geometry.
These  contributions can precisely be matched
on the Type I side.  However, on the Type I side there
are also non-vanishing contributions from poly-instantons,
i.e. from  E1-instantons not lying on top of each other
in the sense that they carry different Wilson lines.  
These corrections are absent on the heterotic side.
The resolution of this puzzle is still an open issue.
If S-duality holds, two options seem conceivable. First, the naive (tree-level)
S-duality map might receive  instanton corrections. Second,
since the starting point for computing the heterotic threshold corrections
involves just the partition function of a single heterotic string,
one might be missing these multiple (poly) worldsheet instanton
corrections from the very beginning. More data seems
to be necessary to settle this issue.\footnote{The heterotic worldsheet instanton corrections to the gauge kinetic
function for a  Gimon-Polchinski like model have
been computed in \cite{cd08}. Here the Type I S-dual
computation still remains to be worked out.}

\label{subsecmultiinst}

\subsection{Instantons and background fluxes}
\label{sec_flux}

The details of the instanton calculus are modified in the presence of non-trivial closed string background fluxes. In general such fluxes can introduce couplings in the instanton effective moduli action which provide new ways to integrate out some of the fermionic zero modes. In this manner extra zero modes  may acquire flux dependent effective mass terms and are thus lifted without inducing higher derivative F-terms.

Suppose for simplicity that an instanton $\cal E$ possesses, in addition to the universal modes $x^{\mu}$ and  $\theta^{\alpha}$, two extra zero modes $\ov \psi^{\dot \alpha}$. At this stage we are not specifiying whether these correspond to deformation, Wilson line or extra universal modes. The situation we are interested in occurs when some background flux $G$ induces a mass term of the form
\bea
\label{flux_gen1}
S_{G} = \int {\cal O}_G \, \ov \psi \, \ov \psi
\eea
with a non-vanishing flux-dependent operator  ${\cal O}_G$. 
Integrating out the $\ov \psi$ modes leads to a contribution of the form
\bea
\label{flux_gen2}
\int d^4 x \, d^2 \theta \, d^2 \ov \psi \, e^{-S_{\cal E} - {\cal O}_G \,\ov  \psi \, \ov \psi} = \int d^4x \,  d^2 \theta \, {\cal O}_G \, e^{-S_{\cal E}}.
\eea

In principle the mechanism of equ. (\ref{flux_gen2}) can work for all types of D-brane instantons in orientifolds. Determining the non-vanishing operators of the form (\ref{flux_gen1})
requires precise knowledge of the couplings of the background fluxes to the instanton zero modes. 
These can be derived either within a supergravity approach starting from the superembedding of the worldvolume of the instanton or by a direct CFT computation of the relevant open-closed couplings.
Most efforts in the literature have focused on the flux-induced lifting of deformation zero modes of M5-brane instantons in M/F-theory \cite{ks05,NS05,kkt05,DT07} and of E3-brane instantons in Type IIB orientifolds \cite{gktt04,tt05,EB05,JP05,lrst05,DL07,WS07}.
More subtle is the effect of fluxes on the extra Goldstinos $\ov \tau^{\dot \alpha}$. It has been analysed for E3-brane instantons  in \cite{bcrw07,MB08a} and for fractional E(-1) instantons  in \cite{MB08a,MB08b}. 
The general compatibility of fluxes and instanton contributions to the superpotential is analysed in the abstract and in concrete backgrounds in \cite{kt05} and \cite{bk07}.

\subsubsection*{E3-instantons in Type IIB orientifolds with 3-form flux}

The probably best understood class of flux compactifications is that of Type IIB Calabi-Yau orientifolds with O3/O7-branes and background 3-form flux. 
We will only consider  compactifications of the type \cite{gkp01} with a non-trivial background value of $ G_3 = F_3 - \tau H_3 $ of RR- and NS-flux $ F_3= d C_{2}$ and $
H_3= dB_2$ and constant dilaton $\tau = C_0 + i e^{-\phi}$. 

To set the stage recall from \cite{gkp01} that for compactifications with D3-, anti-D3 and D7-branes as well as O3/O7-planes as local sorces, the global consistency conditions of the supergravity equations force the 3-form flux to be imaginary self-dual (ISD).  Such ISD flux can be of Hodge type (0,3), (2,1) primitive or (1,2) non-primitive. Primitivity, which amounts to the condition $J \wedge G =0$,  is automatically satisfied for 3-form flux on Calabi-Yau spaces, which do not have non-trivial 5-cycles, but is a non-vacuous  constraint on $T^6$ or $K3 \times T^2/{\mathbb Z}_2$. As shown in \cite{gp00} in absence of any non-perturbative effects only the (2,1) primitive part of the 3-form flux satisfies ${\cal N}=1$ supersymmetry. 
These results were generalised in \cite{lrss05,lrst05,DL07}: In the presence of non-perturbative contributions to the superpotential beyond the flux-induced superpotential \footnote{Here we mean superpotential contributions by instantons which would exist already without taking into account the flux induced lifting of zero modes.}, supersymmetric vacua exist also for flux of Hodge type (3,0), (1,2) and (0,3). This is because the F-terms induced by these fluxes can be cancelled against the F-terms of the dilaton, the complex structure and the K\"ahler moduli, respectively, such that $D_{i} W_{flux} + D_i W_{non-pert.} = 0$.   

Our aim is to determine the flux induced mass terms of the form (\ref{flux_gen1}) for an E3-brane instanton wrapping a holomorphic divisor $\Gamma$ of the Calabi-Yau. 
The result will be given schematically in equs. (\ref{fluxaction_E3}) and (\ref{F_linear}).
For explicitness we focus on the case of an isolated $U(1)$ instanton.\footnote{Recall that isolated $U(1)$ instantons are instantons not invariant under the orientifold action and which do not intersect their orientifold image. In particular there are no ${\cal E}-{\cal E}'$ zero modes.} In general such instantons can carry non-vanishing worldvolume flux ${\cal F}= F- B|_{\Gamma}$, where $F$ is the gauge flux associated with the $U(1)$ gauge group of the instanton. For the time being, however, let us set ${\cal F} =0$. As derived in \cite{mms03,mrbp05,EB05} with supergravity methods and confirmed by the CFT analysis of \cite{MB08a} the 3-form flux couplings in absence of gauge flux take the form 
\bea
\label{Fluxaction1}
S= \int_{\Gamma} d^4 \zeta\sqrt{ {\rm det}g}\, \, \omega \, \,  \left(e^{-\phi} \, \Gamma^{\tilde m} \nabla_{\tilde m} + \frac{1}{8} \, \widetilde G_{\tilde m \tilde n p} \, \Gamma^{\tilde m \tilde n p} \right) \omega.
\eea
Here the 3-form flux appears in the combination  
\bea
\widetilde G_{\tilde m \tilde n p}= e^{-\phi} H_{\tilde m \tilde n p} + i F_{\tilde m \tilde n p}' \gamma_5
\eea
in terms of  $F_{\tilde m \tilde n p}'= F_{\tilde m \tilde n p} - C_0 H_{\tilde m \tilde n p}$ and the four-dimensional matrix $\gamma_5$. The indices $\tilde m,\tilde n$ are along the four-cycle $\Gamma$ and $p$ is transverse to it.

The above action uses a ten-dimensional notation for the fermionic degrees of freedom encoded in the object $\omega$.
Locally $\omega$ can be decomposed into a  four-dimensional chiral (anti-chiral) Weyl-spinor times an internal part $\epsilon_+$ ($\epsilon_-$) given by 
\bea
\label{spinor_decomp}
\epsilon_+ &=& \phi |\Omega\rangle +  \phi_{\ov a}\Gamma^{\ov a} |\Omega\rangle + \phi_{\ov a \ov b}\Gamma^{\ov a \ov b} |\Omega\rangle, \nonumber \\
\epsilon_- &=& \phi_{\ov z} \Gamma^{\ov z}|\Omega\rangle +  \phi_{\ov a\ov z}\Gamma^{\ov a\ov z} |\Omega\rangle + \phi_{\ov a \ov b\ov z}\Gamma^{\ov a \ov b\ov z} |\Omega\rangle.
\eea
This decomposition makes use of the local choice of complex coordinates $a,b = 1,2$ along $\Gamma$ and $z,\ov z$ for the transverse direction as well as the standard definition of the Clifford vacuum
$|\Omega\rangle$,
\bea
\Gamma^z |\Omega\rangle =0, \quad \Gamma^a |\Omega\rangle =0.
\eea

Consider now the flux-induced lifting of what would be a zero mode in the absence of any three-form flux.\footnote{A more general treatment in terms of the full Dirac equation involving a flux-induced torsion piece can be found in \cite{kkt05,EB05}} For $G=0$ the zero modes, i.e. the solutions to the ordinary Dirac equation, are given by the harmonic piece of the modes (\ref{spinor_decomp}).
The universal fermionic zero modes with four-dimensional polarisation $\theta^{\alpha}$ and  ${\overline \tau}^{\dot\alpha}$ can be identified with
\bea
\label{universal_decomp}
\omega_0^{(1)} = \theta^{\alpha} \otimes  \phi |\Omega\rangle, \quad\quad\quad  \omega_0^{(2)} = \overline \tau^{\dot \alpha} \otimes  \phi_{\ov a \ov b\ov z}\Gamma^{\ov a \ov b\ov z} |\Omega\rangle.
\eea
Recall furthermore from table \ref{tabledeform} that the Wilson line and deformation modes are counted by $H^{(0,1)}(\Gamma)$ and $H^{(0,2)}(\Gamma)$, respectively.  
It follows that the Wilson line modulini correspond to $\gamma^{\alpha} \otimes \phi_{\ov a}\Gamma^{\ov a} |\Omega\rangle$ and their conjugates $\ov \gamma^{\dot \alpha} \otimes \phi_{\ov az}\Gamma^{\ov az} |\Omega\rangle$, while the deformation modulini are given by $\chi^{\alpha} \otimes \phi_{\ov a \ov b}\Gamma^{\ov a \ov b} |\Omega\rangle $ (plus their conjugates $\ov \chi^{\dot \alpha} \otimes \phi_{\ov z}\Gamma^{\ov z} |\Omega\rangle $ ).\footnote{Since we are considering here a $U(1)$ instanton away from the orientifold plane the deformation and Wilson line modulini are not subject to the projections of table \ref{tabledeform}. }

Given a particular combination of 3-form flux $G_3$ one can now evaluate its induced couplings to the fermionic zero modes using the action (\ref{Fluxaction1}) and the decomposition (\ref{spinor_decomp}). 
The result is that 
\begin{itemize}
\item
(2,1) flux can in principle couple to the anti-chiral deformation  and to the anti-chiral Wilson line modulini of unmagnetised E3-brane instantons; under appropriate circumstances these can therefore be lifted \cite{tt05,EB05}.
Lifting their chiral counterparts requires (1,2) flux.  
\item primitive (2,1) flux does not couple to  the extra Goldstinos $\ov \tau^{\dot \alpha}$ of unmagnetised E3-brane instantons \cite{bcrw07}; only (3,0) and (2,1) non-primitive flux  couples to these modes. 
Corresponding statements for the $\theta^{\alpha}$ modes hold by conjugation.
\end{itemize}

To illustrate this latter point for (2,1) primitive flux we compute the action e.g. of  ${\widetilde G}_{\ov a b z} \Gamma^{\ov a bz}$ on the internal part of the extra Goldstinos $\omega_0^{(2)}$, $\phi_{\ov a \ov b \ov c}  \Gamma^{\ov a \ov b \ov c} |\Omega \rangle$.  Elementary gamma-matrix algebra reveals that
\bea
\label{tau-coupl_G}
{\widetilde G}_{\ov a b z} \Gamma^{\ov a b z} \Gamma^{\ov 1}  \Gamma^{\ov 2}  \Gamma^{\ov 3 }|\Omega\rangle =
{\widetilde G}_{\ov a b z}  g^{b \ov a} \left(g^{z \ov 1} \Gamma^{\ov 2}  \Gamma^{\ov 3}|\Omega\rangle - g^{z \ov 2} \Gamma^{\ov 1}  \Gamma^{\ov 3}|\Omega\rangle  + g^{z \ov 3} \Gamma^{\ov 1}  \Gamma^{\ov 2}|\Omega\rangle \right)
 = 0.
\eea
The last equation follows from the identity \cite{EB05}
${\widetilde G} |\Omega\rangle = i \, {G} |\Omega\rangle$
together with primitivity of $G$,
$g^{c \ov c'} G_{b c \ov c'} = 0$ \cite{bcrw07}.\footnote{Note for completeness that for non-primitive (2,1) flux the right-hand side of equ. (\ref{tau-coupl_G}) is non-zero and leads to a non-diagonal coupling to some deformation modes $\ov \chi$ of the schematic form $G_{(2,1) {\rm n.p.}} \ov \chi \ov \tau$.}
By the same token one finds that $(3,0)$ flux can lift the $\ov \tau$ modes, while (0,3) and (1,2) flux does not couple to them. These results were confirmed by CFT methods in \cite{MB08a}.

To summarize,  3-form flux on a Calabi-Yau manifold \footnote{Note that in this case no non-primitive (2,1) or (1,2) flux exists.} induces mass terms for unmagnetized E3-instantons of the schematic form
\bea
\label{fluxaction_E3}
S^{E_3}_G &=& \int G_{(0,3)} \, \theta \, \theta + G_{(3,0)} \, \ov \tau \, \ov \tau  \\
&+& \int  G_{(2,1)}^{\rm prim.}\,  \ov \chi \, \ov \chi  + G_{(2,1)}^{\rm prim.} \, \ov \gamma \, \ov \gamma
   + G_{(1,2)}^{\rm prim.}\,  \chi \, \chi  + G_{(1,2)}^{\rm prim.} \, \gamma \, \gamma. \nonumber 
\eea

The lifting of the Goldstinos in  vacua with (2,1) flux requires the interplay of 3-form flux and non-trivial supersymmetric worldvolume flux ${\cal F}$ \cite{bcrw07}.
Indeed, for  ${\cal F} \neq 0$ new interaction terms appear  \cite{tt05}. The BPS condition on the gauge flux amounts to primitivity of ${\cal F}$. 
The part in the instanton effective action relevant for the lifting of the universal modes can be written as  \cite{bcrw07}
\bea
\label{F_linear}
S_G 
\simeq  \,\int_{\Gamma} d^4 \zeta\sqrt{ {\rm det}g} \, \, \omega \, {\cal O}(G_{(2,1)}^{\rm prim.}, {\cal F}) \, \omega \quad\quad {\rm with} \quad
{\cal O}(G, {\cal F}) =  \, {\cal F}_{\widetilde i \widetilde j} \,  \Gamma^{\widetilde i pq} \, g^{ \widetilde j \widetilde k}\,\, G_{\widetilde kpq}.
\eea
Here a tilde denotes indices parallel to the worldvolume, whereas $p,q$ are general internal indices.

In the presence of suitable three-form flux the interaction term (\ref{F_linear})  leads to
a coupling of the zero mode $\omega^{(2)}_0$
proportional to
\bea
\label{foehr}
{G}_{\ov a b z}  {\cal F}^{b \ov a}  g^{z \ov 3} \Gamma^{\ov 1}  \Gamma^{\ov 2}|\Omega\rangle  .
\eea
Note that unlike the coupling (\ref{tau-coupl_G}) this need not vanish by primitivity of ${\cal F}$ and $G$.
In fact, a simple local configuration of a magnetised E3-instanton on a fluxed $T^6/{\mathbb Z}_2$ was found in \cite{bcrw07} where the lifting of the $\tau$-modes via this mechanism can indeed be achieved.

Let us pause a second to interpret these results. The fact that ${\cal N}=1$ supersymmetric background flux alone does not lift the extra Goldstone modulini is surprising, but not inconsistent. After all, their appearance is rooted in the local enhancement of the ${\cal N}=1$ supersymmetry preserved by the orientifold projection to the full ${\cal N}=2$ supersymmetry of the Calabi-Yau compactification away from the orientifold locus. In the presence of background flux this ${\cal N}=2$ supersymmetry is reduced to ${\cal N}=1$ even away from the orientifold plane, and the $\ov \tau$-modes  are no longer protected as the Goldstone modes associated with the breakdown of a global symmetry by the instanton.
Consistently, new interactions can be found which given them a mass term, even though the mere absence of the supersymmetry enhancement is not sufficient for the former Goldstinos to be lifted.

Second, one might wonder if the contribution of an isolated $U(1)$ instanton to the superpotential can be consistent with holomorphicity of the superpotential. This question was analysed in \cite{gmu08}, see also \cite{AU08}. In fact, the isolated instanton considered in the toroidal example of \cite{bcrw07} can become non-supersymmetric across a line of marginal stability upon deforming the K\"ahler moduli as to depart from the primitivity condition $J \wedge {\cal F}=0$. Being an isolated instanton it cannot compensate for the deviation from the BPS condition by  recombination with its orientifold image.  In this case it should not contribute to the superpotential any more. What resolves the paradox in the torodial example is that this deformation of $J$ automatically renders the 3-form flux non-primitive as well. Since in this region of moduli space supersymmetry is broken completely, the non-BPS instanton  need not exhibit additional zero modes which would forbid the generation of a superpotential-like contribution. As of this writing it remains to be seen if the lifting of $\ov \tau$-modes via the coupling (\ref{F_linear}) is available also on genuine Calabi-Yau manifolds.

\subsubsection*{Fractional E(-1)-instantons in Type IIB orientifolds with 3-form flux}

A similar analysis was carried out in \cite{MB08a, MB08b} for fractional E(-1)-instantons at singularities in Type IIB compactifications. These E(-1) instantons can be viewed as E1-brane instantons wrapping a vanshing holomorphic two-cycle.
Unlike in the case discussed above, the relevant flux interactions were analysed here entirely with the help of conformal field theory methods.
For brevity we stick to the case of stringy E(-1) instantons and refer the reader to \cite{MB08b} for a discussion of gauge instantons in the presence of background flux.
Schematically, the 3-form flux couples via \cite{MB08a,MB08b}
\bea
S^{E(-1)}_G \simeq  G_{3,0} \, \theta \, \theta -  G_{0,3} \ov \tau \, \ov \tau.
\eea 
Note that for E(-1) instantons (0,3) flux does indeed couple to the $\ov\tau$-modes. This is in contrast to E3-brane instantons where, as summarised in equ. (\ref{fluxaction_E3}), even (0,3) flux does not couple to the extra Goldstinos.
As a result, it was proposed in \cite{MB08b} (see also \cite{fp09}) that E(-1) instantons of $U(1)$ type can generate superpotential terms in the presence of (0,3) flux.  Of course such instantons are trivially supersymmetric and there arises no paradox from a potential crossing of a line of marginal stability.

\subsubsection*{Interpretation and Outlook}

Some more comments are in order concerning the general philosophy of lifting zero modes by flux-induced terms as in equ. (\ref{flux_gen1}).

Consider first the case of (2,1) primitive flux, which is supersymmetric already by itself.
Suppose the flux lifts some deformation modes of type $(c,\ov \chi^{\dot \alpha})$. As described previously, in absence of flux  these extra zero modes lead to Beasley-Witten multi-fermion interactions.\footnote{As for the deformation modes we are having the anti-chiral deformation modes $(c, \ov \chi^{\dot \alpha})$ in mind. Their chiral counterparts, if not projected out, can only be lifted by  fluxes other than of (2,1) type, see equ. (\ref{fluxaction_E3}). The same applies in principle to the chiral Wilson line modulini $\gamma^{\alpha}$, which are involved in the generation of corrections to the gauge kinetic function. Only the $\ov \gamma^{\dot \alpha}$ can be lifted by (2,1) flux, but they are not relevant for the generation of interesting F-terms.} The effect of the flux induced mass term is therefore to turn the multi-fermion F-terms into contributions to the superpotential.

The resulting generation of a superpotential term can also be understood from a purely four-dimensional effective field theoretic approach. As argued in \cite{AU08} background flux can induce mass terms of the form  (\ref{flux_gen1}) precisely when it also lifts those closed string moduli which would be involved in the Beasley-Witten terms generated by the instanton for vanishing flux. Integrating out these massive closed string moduli indeed transforms the original multi-fermion interactions into a superpotential interaction below the mass scale of the fixed closed string moduli.

Less clear is the lifting of moduli by fluxes of other Hodge type for which a supersymmetric vacuum exists only in presence of a non-perturbative superpotential even before taking the effect of the fluxes into account \cite{lrss05,DL07}.
In \cite{lrst05} it is argued that the naive couplings of this flux to the moduli are cancelled against additional terms which can be understood as the backreaction of the instanton on the setup. It will be interesting to see how the lifting of, say, the $\ov \tau$-modes of E(-1) instantons by (0,3) flux is affected by these considerations.

\subsection{D-terms from non-BPS instantons}
\label{sec_D-term}

While a  thorough discussion of instanton induced D-terms is beyond the scope of this review we would like to briefly point out some  pertinent developments in the recent literature.
Through the lack of holomorphicity instanton corrections to D-terms such as the Fayet-Iliopoulos term or the K\"ahler potential are currently under comparatively poor computational control.
The generation of a D-term requires an instanton whose universal fermionic zero modes span the full ${\cal N}=1$ supersymmetry algebra preserved in four dimensions. In the notation of table \ref{tabGoldstino} these are the modes $\theta^{\alpha},\,  \ov \theta^{\dot \alpha}$. From the general arguments in \S\ref{subsec_zero} we therefore need a non-BPS instanton whose presence indeed breaks all supersymmetries in such a way that the four zero modes $\tau^{\alpha}, \ov \tau^{\dot \alpha}$ are lifted appropriately. 

In \cite{abls07b} it was argued that a Fayet-Iliopoulos D-term can be generated by ${\cal O}(1)$ instantons (with two chiral and two anti-chiral  deformation modulini) which are non-supersymmetric due to a non-zero pull-back of RR potentials to its worldvolume.
Another natural system of non-BPS instantons is given by isolated $U(1)$ instantons which become non-supersymmetric away from the line of marginal stability in hypermultiplet moduli space \cite{gmu08}. In particular this reference discusses how the multi-fermion F-terms generated on the BPS locus pick up proper D-term contributions once the instanton becomes non-supersymmetric. 
Finally, non-perturbative corrections to the  K\"ahler potential by  instanton-anti instanton pairs were considered in \cite{bf08}.

\section{INSTANTONS IN QUIVER THEORIES}
\label{secquiver}
In the discussion to this point, we have implicitly used free worldsheet conformal field theory techniques in quantizing the
open strings that stretch between our stringy instanton and other branes that are present in the background.  However, many
of the most interesting geometries for string compactification are highly curved, giving rise to strongly interacting worldsheet
conformal field theories.  In this section, we describe the existing techniques to infer stringy instanton contributions to
holomorphic couplings even in these situations.  In order to avoid introducing cumbersome notation while still
making the major points
clear, we focus on one particular class of geometries; the generalization of these ideas to other geometries should however
be transparent, and we indicate how it proceeds at various points.  We will mostly follow the
references that have focused on local conifold geometries and their close relatives, but 
other related works with many related results appear in \cite{bk07,abflp07,bfm07,SF07,iu07,afp08,fgu08}.

\subsection{Rules for rigid stringy instantons at singularities}

Starting with the seminal work of Douglas and Moore \cite{Douglas:1996sw}, it has been realized that the
field theories arising on D-branes at
Calabi-Yau singularities can be represented in terms of quiver diagrams.  Here, we assume the reader is familiar with
the basic notions of such diagrams.  Discussions of how to understand the quiver associated to a given geometry can
be found in the reviews \cite{Malyshev:2007zz,Wijnholt:2007vn} and in the many references therein.

\begin{figure}[ht]
  \centering
  \includegraphics[width=5cm]{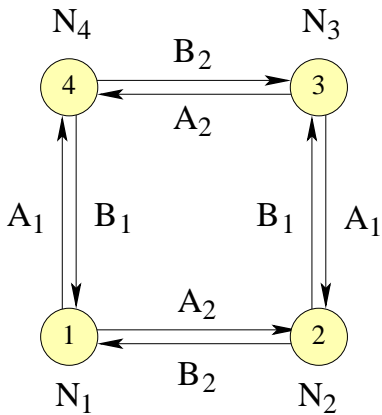}
  \caption{Quiver diagram for the ${\mathbb Z}_2$ orbifold of the conifold,
 for arbitrary numbers of fractional and regular
D3-branes. We have labeled bifundamentals according to the parent field in the un-orbifolded conifold theory.}
  \label{quiver_Z2_conifold_labeled}
\end{figure}

Here, we focus on the quivers arising for D-branes in IIB string theory
probing the singular geometries defined by the constraint
\begin{equation}
(xy)^n = zw
\end{equation}
in $\mathbb C^4$.
These are just ${\mathbb Z}_n$ orbifolds of the conifold; the resulting gauge theories are described in detail in 
\cite{Uranga,ABFKone}.  While the standard conifold quiver theory has two nodes with bi-fundamentals $A_{1,2}$
and $B_{1,2}$ running between them in opposite directions \cite{KW}, the ${\mathbb Z}_n$ quotient gives rise to $2n$ nodes.
The content for $n=2$ appears in  figure \ref{quiver_Z2_conifold_labeled}; the general case is the obvious extension to a larger number
of nodes.  The superpotential governing the matter fields (with the notation that $X_{12}$ is a bifundamental between 
nodes 1 and 2, and $ X_{21}$ is a bifundamental in the conjugate representation) is
\begin{equation}
\label{Wis}
W = h \left( X_{12} X_{23} X_{32} X_{21} - X_{23} X_{34} X_{43} X_{32} + X_{34} X_{41} X_{14} X_{43}
- X_{41} X_{12} X_{21} X_{14} \right)~.
\end{equation}

Because the quiver is completely non-chiral, we are free to occupy the nodes with arbitrary numbers of spacetime filling
branes without inducing any anomalies in the field theory.  Our greatest interest is in stringy instantons, so we focus
on the case where some node is unoccupied.   What is the spectrum of zero modes on such an instanton?

We can follow the general classification of \S\ref{subsecDinst}.    First of all, there are universal zero modes.  Because the
geometry breaks the SUSY to ${\cal N}=2$, while further spacetime filling branes break it to ${\cal N}=1$, the stringy instanton on an unoccupied node will see ${\it four}$ fermionic
zero modes -- two from the 2 broken ${\cal N}=1$ supercharges, and two more from the orthogonal ${\cal N}=1^\prime$.
So a priori, one does not expect any contribution to the superpotential.  This is easily fixed by introducing suitable
orientifold planes into the geometry as in \cite{abfk07,abflp07,bfm07,isu07,bcrw07}.  When we proceed
we will therefore assume that the
instanton wraps a node of the quiver which has an $SO$ orientifold projection on the instanton (and would have an
$SP$ projection on a corresponding spacetime filling D-brane at the same node).  The simplest orientifold action on the
$2n$ node quiver above truncates the number of nodes to $n+1$ (exchanging $2n-2$ nodes pairwise and fixing two of 
them), and places an orientifold (with an $SO$ projection for the instantons) at the two fixed quiver nodes.   It 
changes the ``circular" $2n$-node quiver into a chain with $n+1$ 
nodes, with the orientifold projections on the first and
last node.  For a more detailed discussion, see \cite{AK}.

What about deformation zero modes?  In the quiver gauge theory on spacetime filling branes, the deformation zero
modes shown in table \ref{tabledeform} give rise to adjoint matter fields.  Since the quiver has no adjoints, our stringy instanton will
be ${\it rigid}$.  

Finally, we must classify the charged zero modes.  The arrows in our quiver represent precisely the charged fermionic zero modes 
tabulated in table \ref{tablezero}.  For each arrow entering/leaving the instanton node, one finds a single Grassmann degree of freedom
in the appropriate bi-fundamental representation of the instanton and spacetime filling gauge group.    That is, the chiral
arrows representing chiral matter fields in the quiver geometry for spacetime filling branes, turn into single Grassmann variables
on the instanton world-volume for an instanton occupying the same node.

We therefore see that the instanton zero modes are simply derived from the quiver governing the spacetime filling gauge
theory at the same singularity.  These rules have been discussed in more detail in \cite{Simic,afp08}.  In fact, at the
level of the worldsheet conformal field theory, it is also easy to prove that the interactions of the charged instanton zero
modes with the rest of the quiver fields may be summarized as follows.  One obtains an interaction of the instanton zero
modes with the charged matter for each superpotential coupling which a spacetime filling brane at the same node would have
possessed.  One simply replaces the charged chiral fields entering the instanton node with the appropriate instanton
Grassmann zero modes.

So for instance, in our orientifold above, the charged chiral fields $Q, \tilde Q$ leaving the end-points of the linear quiver
would be replaced by charged instanton zero modes $\lambda, \ov \lambda$.  And they would have an effective
action
\begin{equation}
S_{eff} = t + \lambda X_{23} X_{32} \ov \lambda,
\end{equation} 
where $t$ is the K\"ahler parameter controlling the size of the node the instanton wraps.  In the particular case that 
one has $N_2=1, N_3=N$, for instance, from the integral
\begin{equation}
\int~d\lambda~d\ov\lambda~e^{-S_{eff}} = e^{-t} X_{23} X_{32}
\end{equation}
one infers  that one can generate an exponentially small non-perturbative mass term for the fields $X_{23}, X_{32}$, i.e.
for a flavor of the $SU(N)$ theory at node 3 \cite{abfk07}.   
For $N_2 >1$, one would instead find a higher dimension (irrelevant)
operator in that SUSY QCD theory.  

In summary, the general rules for adding stringy instantons to quiver gauge theories arising at Calabi-Yau singularities are quite simple.  One gets a charged Grassmann zero mode $\lambda$ for each charged chiral field $Q$ emanating from the quiver node
for the related spacetime filling gauge theory.  And, one gets
interaction terms between these Grassmann fields and the matter fields of the space-time gauge theory, for each superpotential
term that would have coupled $Q$ to matter at the other nodes.  

\subsection{Geometric transitions and stringy instantons}
\label{sec_GT}

In some cases, the non-perturbative dynamics on D-branes at a singular geometry $X$ can be determined by performing a 
``geometric transition" to a different geometry $X'$, replacing the branes with fluxes.  In the IIB theory, for instance, there
are three-form fluxes $H_3, F_3$ from the NS and RR sector.  Turning on background values of these fluxes generates
a superpotential for the Calabi-Yau complex moduli \cite{GVW}
\begin{equation}
\label{fluxpot}
W = \int (F - \tau H_3) \wedge \Omega~.
\end{equation}
By appropriate mapping of the D-brane quanta on $X$ to fluxes on $X'$, and the parameters (like the dynamical
scale) of the brane field theory
to the moduli of $X'$, one can determine the non-perturbative dynamics of the QFT by computing (\ref{fluxpot}).

The most famous example of such a correspondence again involves the conifold geometry 
\begin{equation}
\sum_{i=1}^{4} z_i^2 = \epsilon^2~.
\end{equation}
In the limit $\epsilon \to 0$ with real $\epsilon$, 
one can  see that a three-sphere (the real slice of the defining equation above) is
collapsing to zero size.  One can repair this singularity either by deforming the geometry with finite $\epsilon$,
or by performing a small resolution which introduces a finite-sized ${\mathbb P}^1$ at the tip.  The quiver gauge theory
representing the conifold \cite{KW} has two nodes.  Equal occupation of the nodes by $U(N)$ gauge groups describes
N D3 branes on the conifold geometry, while any difference in the numbers $N_2 - N_1 = M$ maps to adding
$M$ additional D5 branes wrapping the small ${\mathbb P}^1$.  

The canonical example of a geometric transition occurs if one has e.g. $N_2 = M, N_1 = 0$ \cite{KS,VafaGT}.
The low-energy gauge theory on the D5-branes is an $SU(M)$ ${\cal N}=1$ pure Yang-Mills theory.
It is expected to have confinement and chiral symmetry breaking at an exponentially small scale
$\Lambda \sim e^{-1/g_{YM}^2}$, with $M$ vacua resulting from a spontaneous breaking of a non-anomalous
${\mathbb Z}_{2M}$ R-symmetry down to ${\mathbb Z}_2$.

In this geometric transition, $X$ is the resolved conifold and $X'$ is the deformed conifold.  The $M$ D5-branes map
to $M$ units of $F_3$ flux through the small sphere $A$ in $X'$, while the NS 3-form flux through the non-compact dual
cycle to this $S^3$, $B$, is chosen to be $t$,  the ${\mathbb P}^1$ volume in X (controlling the gauge theoretic
coupling on the D5s).  The end result is
\begin{equation}
\int_A F_3 = M,~\int_B H_3 = {t },~W = -i {\, t\over g_s} z + M 
\left({z\over 2\pi i}\right) \log(z),
\end{equation}
where $z$ is the size of the $S^3$ (the complex modulus determining $\epsilon$).  
Here, in computing $W$ in terms of $z$, we have used basic facts about the periods of $\Omega$ in the conifold geometry \cite{delaossa}.
Minimizing the flux potential on $X'$, we in fact discover $M$ vacua with
\begin{equation}
\vert z\vert  \sim e^{-2\pi t / g_sM}~,
\end{equation}
precisely in accord with the gauge theory expectations.  The parameter $z$ in the $X'$ geometry represents the dynamical
scale $\Lambda$ in the dual gauge theory.

It is interesting to ask whether one can similarly perform transitions to sum up purely stringy non-perturbative effects, like
the stringy instantons.  Here, following \cite{abk07} (see also \cite{bmv07}), we give an example where in fact a stringy instanton effect (and an
infinite series of multicovers) are reproduced as expected by a geometric transition.  This provides an alternative check
on our computations.

The geometries $X$ for us will be non-compact Calabi-Yau threefolds which are $A_r$ ALE
spaces fibered over the complex $x$-plane.  These are described by hypersurfaces in 
$\mathbb C^4$ via a defining equation
\begin{equation} 
\label{geoms}
uv = \Pi_{i=1}^{r+1} (z - z_i(x))~.
\end{equation}
This geometry is singular at the points where $u, v = 0$ and $z_i(x) = z_j(x) = z$.  At these
points there are vanishing ${\mathbb P}^1$s, which can be blown up by deforming the K\"ahler
parameters of the Calabi-Yau (analogous to the small resolution of the conifold).  There are
$r$ such two-cycle classes, which we will denote by $S_i^2$.  These correspond to the blow-ups
of the singularities at $z_i = z_{i+1}$ for $i=1,..,r$.  By wrapping D5-branes on these spheres,
we can engineer the gauge theories of interest.  The study of
transitions in such geometries was pioneered
in the paper \cite{Cachazo:2001gh}.

If the $z_i$ were independent of $x$, then this geometry would become the product of an $A_r$
ALE space with the $x$-plane.  In this circumstance, D5-branes wrapped on the small ${\mathbb P}^1$s
would have a moduli space; they would have adjoint fields whose vev parameterizes their location
on the $x$-plane.   The superpotential which the non-trivial fibration induces for these adjoints
can be computed as follows \cite{Witten:1997ep,Aganagic:2000gs}.  For the $i$th brane stack,
introduce a 3-chain ${\cal C}$ whose boundary is $S_i^2$.  Then
\begin{equation}
W = \int_{\cal C} \Omega~.
\end{equation}
For this particular geometry, one can show that it simplifies to
\begin{equation}
\label{adjsup}
W_i = \int_{\cal C} \left( z_i(x) - z_{i+1}(x) \right) dx
\end{equation} 
for the $i$th adjoint field.  

In addition to the adjoints, there are quarks stretching between D5-branes wrapped on adjacent
${\mathbb P}^1$s, and they have ${\cal N}=2$ - like couplings to the adjoints.  So the full superpotential is
\begin{equation}
\label{fullsup}
W_{\rm total} = \sum_i  W_i(\Phi_i) + Tr (Q_{i,i+1}\Phi_{i+1} Q_{i+1,i} -
Q_{i+1,i}\Phi_i Q_{i,i+1})~.
\end{equation}
Note that if one chooses the $z_i$ so that the adjoints are all massive with equal vevs, 
then after integrating out
the adjoints, one gets precisely the quiver geometries studied in the previous subsection.
In that sense, this is the simplest generalization of those orbifold geometries.

In fact, we will generalize the previous subsection in another way as well.  While in \S4.1\ we
focused on instantons wrapping empty (orientifolded) nodes, here we will instead sum
up instantons on $U(1)$ nodes.  These are still ``stringy," since the $U(1)$ gauge theory does
not have smooth Yang-Mills instantons.   And as described in \S3.1, the zero mode selection
rules are expected to allow contributions even in this case.  We will perform a geometric transition
on such a $U(1)$ node, and use the dual geometry to show that one gets non-perturbative
generation of an exponentially small mass term (as also in the previous subsection).

We consider the $A_3$ case of (\ref{geoms}), choosing the $z_i(x)$ so that
\begin{equation}
\label{fayet}
uv = (z-mx) (z+mx) (z-mx) (z+m(x-2a))~.
\end{equation}
After blowing up, in our geometry $X$ (pre-transition)
we wrap $M$ branes each on $S_1^2$ at $z_1(x) = z_2(x)$ and $S_2^2$ at $z_2(x) = z_3(x)$, and a single brane on
$S_3^2$ at $z_3(x) = z_4(x)$.
The tree level superpotential is then
\begin{equation}
W = \sum_{i=1}^3 W_i(\Phi_i) + Tr(Q_{12} \Phi_2 Q_{21} - Q_{21} \Phi_1 Q_{12} ) +
Tr(Q_{23}\Phi_3 Q_{32} - Q_{32} \Phi_2 Q_{23})~,
\end{equation}
with the $W_i(\Phi_i)$ taking the values
\begin{equation}
W_1 = m \Phi_1^2,~W_2 = -m \Phi_2^2,~W_3 = m (\Phi_3 -a )^2~.
\end{equation}

The adjoint superpotentials have localized brane stacks 1 and 2 at $x=0$.  So they intersect,
and the intervening quark flavors remain massless after inserting the adjoint vevs.  However,
the third node is localized at $x = a$; both its adjoint and its quark matter are massive, and
hence it is a fully massive node (ignoring the free abelian gauge field).  Correspondingly, we
expect that we should be able to perform a geometric transition on this node.  

The result deforms the (formerly resolved) singularity after shrinking $S_3^2$, changing the
complex structure to that of a new manifold $X'$
\begin{equation}
uv = (z-mx) (z+mx) ((z-mx)(z+m(x-2a)) -s)~.
\end{equation}
The size of the new ``deformed" $S^3$ which replaces $S_3^2$ is
\begin{equation}
\int_{S^3} \Omega = S = {s\over m}~.
\end{equation}

Since the third D5-brane is gone, so are the fields $Q_{23}, Q_{32}$ and $\Phi_3$.  We
replace
them in the effective superpotential instead by the flux superpotential for $S$, and by the
deformed superpotential for $\Phi_2$ (which has had its potential changed since we have
integrated out fields that it couples to).  The result is
\begin{equation}
W_{eff} = W_1(\Phi_1) + \tilde W_2(\Phi_2,S) + Tr (Q_{12}\Phi_2 Q_{21} -
Q_{21} \Phi_1 Q_{12}) + W_{\rm flux}(S)~.
\end{equation}
In this geometry, the exact flux superpotential is as in the case of the conifold that we previously
discussed; the geometric transition is ${\it locally}$ identical.  The new superpotential term
$\tilde W_2$ is
\begin{equation}
\tilde W_2(x) = \int (z_2(x)  - \tilde z_3(x)) dx
\end{equation}
where we define $\tilde z_3(x)$ via the relation
\begin{equation}
(z - \tilde z_3(x)) (z - \tilde z_4(x)) = (z - z_3(x)) (z - z_4(x)) - s 
\end{equation}
with $\tilde z_3$ being chosen on the branch which asymptotes to $z_3(x)$ at large $x$.
Concretely, this implies
\begin{equation}
\tilde W_2(x) = \int_{\Delta}^{x} (- m(x' + a) - \sqrt{m^2 (x'-a)^2 + s}) dx'~.
\end{equation}
Here, $\Delta$ is an IR cut-off in the geometry (which maps to a UV cutoff in the field theory).

One can now integrate out $S$ and the remaining adjoints.  It is intuitively quite clear that
the superpotential $W_{\rm flux}$ stabilizes $S$ at an exponentially small value
$S \sim e^{-t/g_s}$.  Then, plugging the vev of $S$ into $\tilde W_2$, and expanding
in powers of $s = mS$, one finds an infinite series of instanton contributions.  Their leading effect is to shift the vacuum for $\Phi_2$ a bit away from its old location by an amount of order $e^{-t/g_s}$;
so one obtains from the term
\begin{equation}
W_{eff} = ...+ Q_{21} \Phi_2 Q_{12} + ... \to e^{-t/g_s} Q_{12}Q_{21}~.
\end{equation}

In other words, one obtains an exponentially small mass for the quarks stretching between the
remaining nodes.  This is of course reminiscent of the phenomena described in \S4.1.
The precise formulae, including small corrections to the above scalings and coefficients for
all terms in the multi-instanton series, can be found in \cite{abk07}.

\subsection{Another check: stringy instantons in RG cascades}
\label{sec_RG}

One of the most interesting phenomena discovered in quiver gauge theories is RG cascades,
where as one moves towards the infrared, the effective gauge theory description changes by a
self-similar sequence of Seiberg dualities.  The simplest example occurs in the conifold quiver
with unequal ranks \cite{KS}; generalizations to the orbifolded conifold geometries of \S4.1
are also easy to exhibit \cite{ABFKone}.

We argued in \S4.1\ that if one orientifolds the geometry $(xy)^n = zw$, one can obtain
a quiver with $n+1$ nodes.   The gauge theory realized by spacetime filling branes in this
quiver is $Sp(N_1)\times U(N_2) \times...\times U(N_n) \times Sp(N_{n+1})$ -- i.e., there is
an $SO$ projection for Euclidean branes on the first and last nodes.  

We also argued that in the special case with e.g. $N_1 = 0, N_2 = 1, N_3 = N, ...$, one obtains
from a stringy instanton on the first node a non-perturbatively generated mass for the quarks
$X_{23}, X_{32}$ stretching between nodes 2 and 3. 

The existence of the RG cascade offers us another possibility to check this claim \cite{AK}.
Suppose we start with large occupation numbers at the nodes, chosen so that at the final step
of the RG cascade, we end up with the configuration above.  Then, one should be able to
derive the effective low-energy theory in two different ways:

\noindent
1) One can do the path integral over D-instanton zero modes at node 1, with occupation
numbers in the quiver gauge theory describing the final cascade step.  This is the class of
techniques we have been describing in this review.

\noindent
2) One could also try to derive the ${\it same}$ effective low-energy theory by analyzing the gauge
theory at higher steps in the cascade, where the relevant node is occupied by spacetime filling branes.
In this case, one should be able to reproduce the ``stringy instanton'' effect by using standard
techniques and results in ${\cal N}=1$ supersymmetric gauge theory.  

While going through the details of the renormalization group cascades for orientifolded orbifolded
conifolds is beyond the scope of this brief review, we simply state here the results.  The gauge theory
analysis is completely consistent with the microscopic expectation from instanton calculus:
in the case that one has the orientifolded quiver, with configuration $N_1 = 0, N_2 = 1, N_3 = N,....$
at the final cascade step, one can prove from gauge theory analysis that gauge dynamics 
generates an exponentially small mass for the quarks $X_{23}, X_{32}$.  In the case that one
studies the cascade with only $U(N)$ nodes and no orientifold, one does not generate such a mass
via gauge dynamics in the cascade.  It is interesting that in the case that the non-trivial effect occurs,
the stringy instanton effect turns into a strong coupling effect (and not an instanton effect) in the
cascading gauge theory \cite{AK}.
Extensions of these results, giving more cases where alternative UV completions involving gauge theory can be used to derive
stringy instanton effects, have also been noted in \cite{gu07,DK08,agm08}.\footnote{An interesting relation between stringy instantons and matrix models is observed in \cite{IG08} (see also \cite{fp09}). A possible interpretation of some stringy instantons as octonionic field theory instantons is considered in \cite{MB09}. }

We emphasize here, however, that the UV completions involving small numbers of D-branes in a Calabi-Yau geometry, or larger rank non-Abelian theories which produce strong dynamics, are ${\it different}$.  
These results should not be interpreted as indicating that the effects are not stringy; rather that in 
cases where a duality relates the stringy configuration to a (UV distinct but IR equivalent) field theory,
the results of computations are in accord as expected.

\section{PHENOMENOLOGICAL IMPLICATIONS}
\label{sphenoappl}

In this section we highlight specific phenomenological implications of D-brane instanton generated couplings for string model building. Our main focus will be on
superpotential  corrections in the  charged matter sector. At the end of this section 
 we will also  comment on some implications of other types of coupling corrections.
A general overview of the status of F-term corrections due to (multi-) instantons  has been provided in \S\ref{sGenerationFterms}.


As discussed in \S\ref{sBasics} and \S\ref{sGenerationFterms}, an instanton configuration can contribute to the superpotential provided all uncharged fermionic zero modes other than the two universal $\theta^{\alpha}$ are lifted or saturated without inducing higher derivative terms. In particular rigid $O(1)$ instantons are natural candidates to generate  superpotential 
corrections. 
Recall from \S\ref{subsec_zero}   that such an object
has  only two uncharged fermionic zero modes $\theta^\alpha$ to begin with, which,  along with four  bosonic ones $x^\mu$, constitute the universal superpotential zero mode measure $\int d^4x\, d^2\theta$. 

In the presence of charged zero modes $\lambda_{{\cal E}a_i}$ (see table \ref{tablezero}) the superpotential involves charged matter fields, and can thus have drastic effects for string model building.
 The form of the matter couplings is such that the total $U(1)_a$  charges of the zero modes are canceled by those  of the charged matter superpotential terms, see equ. (\ref{instcharge}).

A detailed analysis of the superpotential calculus is spelled out in \S\ref{subsecsuperpot} where the  holomorphic superpotential couplings  (\ref{superpotex}) are  determined explicitly in terms of the  classical instanton action, disc diagram couplings of matter fields with two charged zero modes and 
the holomorphic part of the annulus diagram (see figure \ref{figannu}). 
 Such a superpotential coupling  of mass-dimension D is thus of the form
\be
\mu^D= x M_{s}^D  \exp (-{S_{\cal E}^{(0)}})\, , \label{massD}
\ee
where  we have introduced the string mass scale  $M_{s}=\ell_s^{-1}$.
The tree-level string action $S_{\cal E}^{(0)}$ is determined by the volume in string mass 
units of the cycle wrapped by the instanton as in equ. (\ref{Szero}). The pre-factor $x={\cal O}(1)$ 
 can be determined precisely by carrying out the disc and annulus diagram 
 calculations  when a conformal field theory description of the 
string model is available.

It is of utmost importance that the instanton suppression factor does in general not
coincide with ${8 \pi}/{g_{YM}^2}$ as would be the case for a gauge instanton,
which is given by a Euclidean brane wrapped along the same cycle as a 
matter brane. 
For example in the Type IIA framework of intersecting D6-branes  
the exponent of the classical  E2-instanton action can be cast into the form
\bea
\label{ratioA}
S_{\cal E}^{(0)}=\frac{8\pi^2}{g_a^2}\frac{ 
\hbox{Vol}_{E2}}{\hbox{Vol}_{D6_a}}\, ,
\eea
where $\hbox{Vol}_{E2}$  and $\hbox{Vol}_{D6_a}$ are the respective  volumes 
of 
the three-cycles wrapped by the E2-instanton and D6$_a$-branes in the internal space, and $g_a$ is the  gauge coupling of the $U(N_a)$ gauge theory on  the D6$_a$-branes. These couplings introduce a new hierarchy: in appropriate circumstances the ratios of the volumes of the instanton and D-brane can be just right to generate the  desired magnitude of specific non-perturbatively induced couplings.
In this context we stress that given a particular string vacuum the instanton effects cannot be turned on or off at will, but are determined by the internal geometry. In particular the modulus describing the  ratio of cycle volumes as in equ. (\ref{ratioA}) is constrained by the standard D-term supersymmetry conditions for the $U(1)$ gauge fields living on the spacetime-filling D-branes. See e.g. the reviews \cite{bcls05,dk06,bkls06,AU07,FM07,DL08} for more information.

In the sequel we  shall highlight 
some phenomenological implications of  these  superpotential couplings. 
We focus on mass-dimension  three, two, one and zero couplings,  
where effects could be significant, and mention some potential effects 
for non-renormalizable operators.

\subsection{Mass dimension three couplings}

In absence of any charged zero modes the contribution to the superpotential of
 an ${\cal O}(1)$ D-brane instanton is a function of the closed string moduli via the dependence (\ref{Szero}) of the tree-level suppression factor $S_{\cal E}^{(0)}$ on the instanton volume. For example in Type IIB compactifications this yields an exponential dependence of the superpotential on the K\"ahler moduli and the dilaton.  As the perturbative flux-induced superpotential only involves the complex structure, but not the K\"ahler moduli, E3-instantons are therefore vital in attempts to stabilize the moduli in Type IIB orientifolds with fluxes. This was pioneered in \cite{kklt03} 
and demonstrated in very explicit examples e.g. in \cite{ddf04,ddfgk05,lrss05,DL07}.  
 A modified scenario where  these E3-instanton contributions to the superpotential are balanced against perturbative corrections to the K\"ahler potential was developed in \cite{bbcq05}, while its mirror dual Type IIA construction was worked out in \cite{ptw08}.
On the other hand, the complex structure dependent one-loop Pfaffian $f^{(1)}_{\cal E}$ of E3-brane instantons is quite difficult to extract in concrete settings. In general it depends also on the open string moduli of other D-branes and can become relevant for their stabilization, as e.g. for D3-branes \cite{DB06}. 
Of particular use in generating hierarchies in the closed string moduli sector can be the double suppression by the poly-instanton effects described around equ. (\ref{tower1}) \cite{bmp08}. 
In all these setups it is crucial that the instanton does not intersect any other D-brane to avoid chiral charged zero modes. In particular the volume modulus associated with cycles wrapped by chirally intersecting D-branes cannot be stabilized in this manner \cite{bmp07}.

\subsection{Mass dimension two couplings}

\begin{figure}[ht]
\centering
\includegraphics[width=0.5\textwidth]{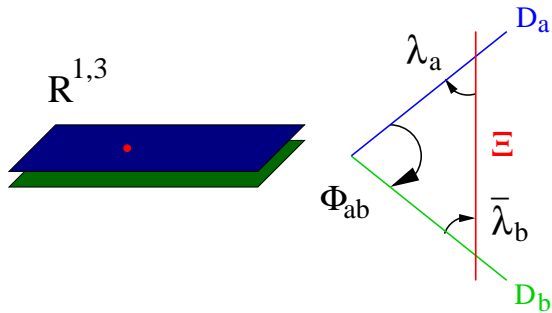}
\caption{The disc diagram for a  Polonyi-type coupling, represented in the Type IIA framework.}
\label{linear}
\end{figure}

Linear couplings can be generated non-perturbatively for matter fields $\Phi_{ab}$ which are charged  under (massive)  Abelian  factors $U(1)_a\times U(1)_b$ only. Namely,
the  charge condition (\ref{instcharge})   for fermionic zero 
modes in this case  requires  $N_a=N_b=1$, along with   one   
$\lambda_{{\cal E}a}$ and one $\lambda_{b{\cal E}}$ mode. For $\Phi_{ab}$  in representation $(-1_a,1_b)$, the  $O(1)$ instanton  must have the following non-zero topological intersection numbers,
 \bea
 \label{int_P}
 I_{{\cal E},{\cal D}_a}^+=1\,  , \ \ \  I_{{\cal E},{\cal D}_b}^-=1\, . 
 \eea
A single  disc  diagram   generates the effective instanton action term $\lambda_{{\cal E}a}\Phi_{ab}\lambda_{b{\cal E}}$. After   absorption  of the fermionic zero modes
 this generates a superpotential term linear in $\Phi_{ab}$. This is illustrated in figure \ref{linear} for an E2-instanton along the cycle $\Xi$ in the framework of Type IIA models with intersecting D6-branes.  
A particular challenge in implementing this and similar effects in phenomenologically appealing string models is associated with the requirement that the instanton may have intersections only with those D-branes which host the respective matter fields in the bifundamental sector. Any additional charged zero mode beyond (\ref{int_P}) will lead to higher dimensional couplings at best and thus annihilate the effect.

\subsubsection*{Supersymmetry breaking}

These mass-dimension two couplings W = $\mu^2 \Phi$  can trigger 
F-term supersymmetry breaking {\`a} la Polonyi 
at the scale $\mu^2$.
Originally \cite{Aharony:2007db} this scenario was envisaged in the special case where the Polonyi field $\Phi$ arises from the ${\cal D}_a - {\cal D}_a$ sector of a single D-brane rather than at the intersection of two branes ${\cal D}_a, {\cal D}_b$. Such a setup requires a vectorlike pair of zero modes $\lambda_{a{\cal E}}, \lambda_{{\cal E}a}$. 
In both variants D-brane instantons can account  not only for the presence of such a  Polonyi term 
\cite{fkms06,Aharony:2007db} but also give an appealing explanation of the hierarchical suppression of its scale $\mu$. This was demonstrated  even in globally consistent examples in \cite{cw07,cw08} based on chiral $SU(5)$ GUT constructions of Type I theory and in \cite{acdl09} on the orbifold $T^6/({\mathbb Z}_2 \times {\mathbb Z})$ with torsion.

This Polonyi-type supersymmetry breaking can in principle be embedded into a scenario of gauge mediation 
 \cite{gr98}  via  perturbative Yukawa couplings of the type $\Phi_{ab} M_{bc} M_{ca}$. Here it is understood that the Standard Model gauge symmetry is part of the gauge group factor $U(N_c)$. In this case the fields $M_{bc}, M_{ca}$ play the role of messenger fields. This scenario has been investigated in various contexts in \cite{abfk07,bf08,cw08,hmssv08,mss08a,mss08b,hv08} (see also \cite{JK07}).

\subsubsection*{Next to the Minimal Supersymmetric Standard Model}

The field $\Phi_{ab}$ can also play the role of a Standard Model singlet field in the next to the minimal supersymmetric Standard Model (NMSSM). Its
perturbative coupling to the Standard Model Higgs doublets $H_{bc}$ and $H_{ca}$ can induce the  $\mu$-parameter after $\Phi_{ab}$ acquires a non-zero vacuum expectation value in the desired regime, triggered by  the D-instanton induced linear couplings for  $\Phi_{ab}$ and supersymmetry breaking in a separate,  ``hidden'' sector.
Further investigation of these types of models is underway \cite{cl09} (see also \cite{hv08}). 


\subsection{Mass dimension one couplings}


\subsubsection*{Neutrino Majorana masses}

\begin{figure}[ht]
\centering
\includegraphics[width=0.5\textwidth]{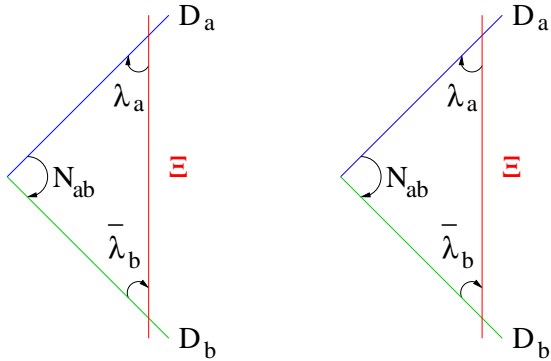}
\caption{Two  disc diagrams  contributing to the  Majorana mass term.}
\label{quadratic}
\end{figure}

Perhaps the most prominent example of non-perturbatively generated mass terms 
are Majorana masses for right-handed neutrinos \cite{bcw06,iu06}.
The prototype example involves fields $\Phi_{a b}$  which are singlets under the Standard Model gauge symmetry, and are charged under additional massive Abelian gauge group factors $U(1)_a\times U(1)_b$, say in the $(-1_a, 1_b)$ representation.
 The condition on the  topological instanton intersection numbers
\bea
\label{int_Maj}
I_{{\cal E},{\cal D}_a}^+=2\, , \ \ I_{{\cal E},{\cal D}_b}^-=2\, 
\eea
then ensures that two disc diagrams (depicted schematically in the Type IIA framework in figure  \ref{quadratic})  generate
superpotential terms quadratic  in $\Phi_{ab}$.
For the desired magnitude of Majorana masses  in the  range $\mu_1={\cal O}(10^{10}-
10^{15})$ GeV, the volume of the instanton cycle has to lie in the
corresponding regime. Furthermore, the constructions should allow  for  perturbative Dirac neutrino Yukawa couplings $m_D$ of the order of the charged sector of the Standard Model $m_D={\cal O} (0.1-10)$ GeV. This results in a see-saw mechanism with  physical neutrino masses $\sim m_D^2/\mu_1$ of the order of $(10^{-2}-10^{-3})$ eV.

This effect was realized within a locally defined chiral GUT theory  with gauge symmetry $U(5)_{GUT}\times U(1)_a\times U(1)_b$ in the Type IIA framework  of intersecting D6-branes on an orientifold of $T^6/({\mathbb Z}_2\times {\mathbb Z}_2')$. In \cite{crw07} a class of ${\cal O}(1)$ instantons with precisely the instanton numbers (\ref{int_Maj}) is identified   that can produce the desired hierarchy for Majorana masses. The explicit 
conformal field theory on such orientifolds 
also allows for an explicit calculation of both the disc \cite{crw07} 
and annulus \cite{abls07b} contributions to  $\mu_1$. 
A systematic search  for globally consistent three-family Standard models with D-instanton induced Majorana masses was performed in \cite{isu07} for models based on Type IIA rational conformal field theories.
The first globally consistent examples with the desired hierarchy for Majorana masses 
were presented in the context of chiral Type I models on elliptically fibered Calabi-Yau spaces in \cite{cw07}.
Further phenomenological implications  of D-brane instanton induced neutrino masses including the possible generation of a realistic family structure were studied in \cite{aim07}.

\subsubsection*{$\mu$-parameter}
A potential explanation of the hierarchically small $\mu$-term $\mu H^u H^d$ can involve D-brane instantons in situations where a perturbative $\mu$-term is forbidden explicitly by the global $U(1)$ charges of the $H^u, H^d$ fields \cite{bcw06,iu06,bmmvw06}. 
For typical  Standard Model  constructions Higgs doublets arise from a chiral sector, say   $H_{ac}$ and $H_{cb}$  in respective representations $(-1_a,{\bf { {2_c}}})$ and $({\bf \overline  2_c}, 1_b)$ under  $U(1)_a\times U(1)_b\times U(2)_c $
($SU(2)_L\in U(2)_c$).  An instanton with the  non-zero topological intersection numbers (\ref{int_P})  has the correct zero mode structure to generate the $\mu$-term due to the quartic  disc diagram  coupling  $\lambda_{{\cal E}a} H_{ac} H_{cb}  \lambda_{b \cal E}$  in the effective  instanton action.
In this case the classical instanton action requires a stronger 
suppression than for Majorana neutrino masses in order to achieve 
$\mu={\cal O}(\hbox{TeV})$. A concrete realisation appears e.g. in \cite{iu07}.

\subsubsection*{Decoupling of non-chiral exotics and further effects}
Many explicit string models with intersecting D-branes are plagued by the appearance of unwanted exotic matter fields.
For example non-chiral matter exotics in the (anti-)symmetric representation of 
$U(N_a)$  arise due to non-zero topological intersection numbers $I_{{\cal 
D}_a,{\cal D}_a\prime}$. It turns out that rigid $O(1)$ instantons with
appropriate intersection numbers can  ensure the correct number of
charge instanton zero modes and  generate mass terms for these non-chiral
exotics. This mechanism has been demonstrated within globally consistent
models both on the Type I \cite{bk07}  and the Type IIA \cite{bclrw07} side.

As yet another interesting implication of instanton generated mass terms, \cite{acdl09} investigates the so-induced breakdown of perturbatively realised conformal invariance in a globally consistent toroidal orbifold.

\subsection{Mass dimension zero couplings}
Both  $SU(5)$ GUT and  multi-stack Standard Model constructions 
in the  Type II framework  generically suffer from the absence of certain desired Yukawa couplings. 
 D-brane instantons  are natural candidates to generate such
perturbatively absent terms.  On the other hand, in typical multi-stack Standard-Model like constructions D-instantons could in principle generate R-parity violating couplings; these in turn would induce experimentally excluded lepton and baryon violating processes. The absence of such dangerous interactions even at a non-perturbative level might thus further constrain the model building possibilities.

In the sequel we exemplify the implications for different tri-linear couplings. 
 We also refer the  reader to \cite{ir08} for a 
systematic analysis of 
$O(1)$ instanton generated  tri-linear couplings (as well as mass terms) 
for  specific four-stack quiver Standard Model constructions.
One should point out that a given instanton can induce more than one 
couplings. Since in this case all associated 
non-perturbative couplings are suppressed by the same classical 
instanton action, it can happen that some couplings turn out to be just of the 
desired magnitude while others may then be too large.
Examples of this phenomenon were also encountered in \cite{ir08}.

\subsubsection*{Top quark Yukawa couplings in SU(5) GUT's}
\begin{figure}[ht]
\centering
\includegraphics[width=0.8\textwidth]{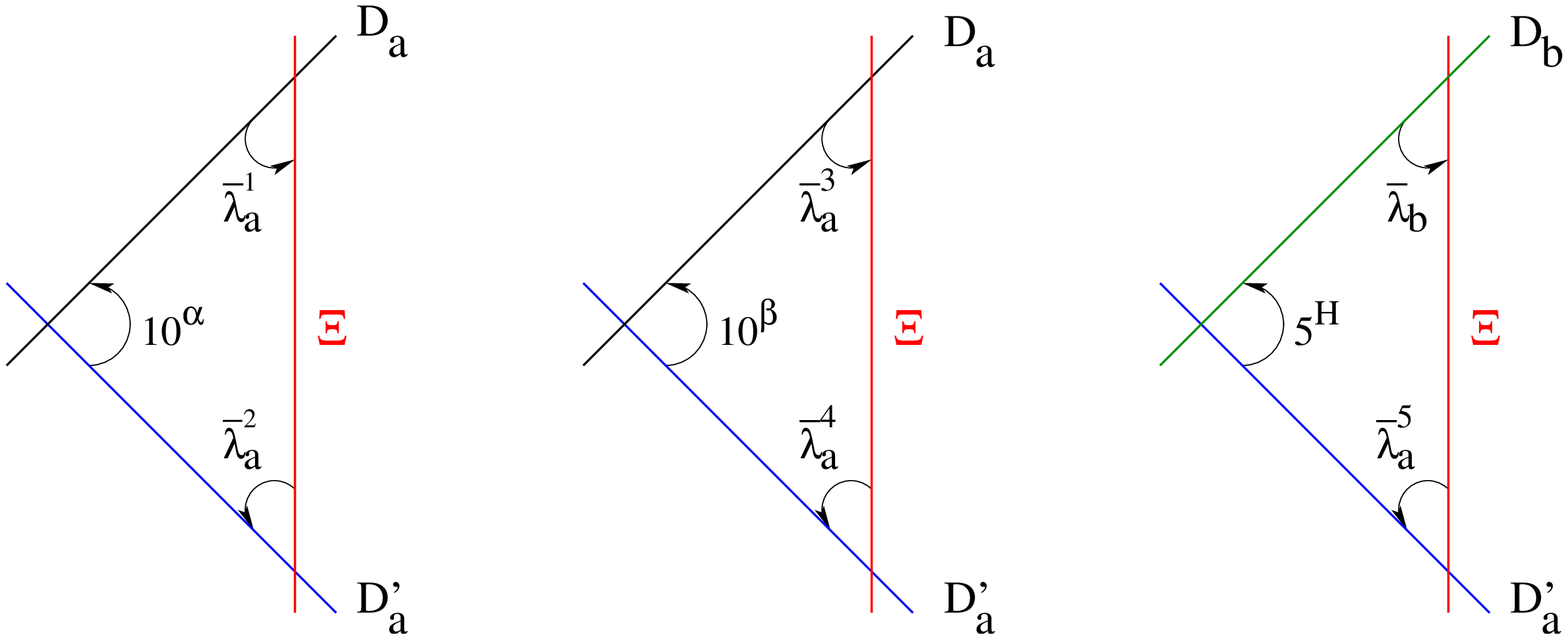}
\caption{Three  disc diagrams contributing to the  top quark Yukawa coupling.}
\label{trilinear}
\end{figure}

Perhaps the most glaring deficiency of $SU(5)$ GUT constructions in Type II orientifolds is the absence of a perturbative top quark Yukawa coupling.
This coupling is of the  type ${\bf 10}^{(2,0)} \, {\bf 10}^{(2,0)} \, {\bf { 5}}^{(1,1)}_{ H}$. 
Here the superscripts denote the $U(1)_a \times U(1)_b$ charges of the respective fields in a minimal two-stack SU(5) setup based on gauge groups $U(5)_a \times U(1)_b$ (where the diagonal Abelian factors are assumed to acquire St\"uckelberg masses via the Green-Schwarz mechanism). 
Evidently this coupling is not invariant under the $U(1)$ charges.\footnote{Note that within non-perturbative F-theory constructions \cite{bhv08b,dw08} these couplings are in principle allowed.} It was shown in \cite{bclrw07} 
that a rigid  $O(1)$ instanton can generate this coupling  provided 
its  non-zero intersection numbers are
\be
 I_{{\cal E},{\cal D}_a}^-= I_{{\cal E},{\cal D}_b}^-=1\, .
 \ee
 In this case the three  disc diagrams illustrated in figure  \ref{trilinear}  generate the top Yukawa coupling.
The desired  hierarchy for this Yukawa coupling requires an appropriately small volume of the instanton cycle. This was achieved explicitly in the globally consistent $SU(5)$ GUT models constructed in \cite{bbgw08} on Type IIB orientifolds with D3/D7-branes.\footnote{An explicit globally consistent realisation of the coupling ${\bf {15}}\,{ \bf {15}}\, {\bf {15}}$ based on gauge group $U(6)$, from which the  ${\bf 10}\,{\bf 10}\, {\bf 5}$ emanates upon brealing $U(6) \rightarrow U(5) \times U(1)$, appears in \cite{iu07}.} Note, however, that the instanton cycle has to be of   string scale size. While one might be worried that one therefore has to sum up the infinite series of all multiply-wrapped instanton corrections this is actually not the case: $N$-fold wrapped instantons along the small cycle cannot contribute to this Yukawa coupling but only generate string-scale suppressed higher dimensional operators as the charged zero mode sector comprises $N$ times more modes. 
By contrast, in flipped $SU(5)$ models the ${\bf 10}^{(2,0)} \, {\bf 10}^{(2,0)} \, {\bf { 5}}^{(1,1)}_{ H}$ coupling accounts for the mass of the down-type quark and can more naturally produce the associated hierarchy. 
\subsubsection*{Dirac neutrino masses}
Another interesting framework to explain the smallness of neutrino masses is given by models where the Dirac neutrino masses are absent perturbatively. 
This can occur if the  anomalous $U(1)$ charges of the right-handed neutrino do not allow for such couplings at a  perturbative level. Appropriate $O(1)$ instanton intersection numbers may in turn ensure the non-perturbative appearance of such couplings. For details  and a concrete local Type IIA construction based on a chiral $SU(5)$ GUT  see   \cite{cl08}. In this case
the hierarchical coupling emerges naturally without further tuning of the volume of the instanton cycles. Namely, for $S_E \sim 8\pi^2/g_{GUT}^2$ the Dirac neutrino masses are of the order of $10^{-3}$ eV.

  \subsubsection*{R-parity violating couplings}
In multi-stack constructions R-parity violating couplings are absent 
perturbatively. However, it was shown that $O(1)$ instantons can generate 
both baryon number violating tri-linear couplings \cite{bcw06,isu07} as well as lepton number violating ones \cite{isu07}. There exist  strong experimental limits on such couplings; in particular  strong bounds from proton decay basically exclude the existence of both types of couplings at the same time. This example shows how important detailed knowledge of the non-perturbative sector of a vacuum can be.

\subsection{Negative mass dimension  couplings}
Non-renormalizable terms induced by instantons are typically subleading since 
they are not only exponentially suppressed by the instanton effective action, but in addition by
powers of the inverse string mass scale.
Nevertheless, such terms could in principle introduce left-handed neutrino masses via  Weinberg operators of the type $(LH_u)(LH_u)/\mu$ \cite{isu07}. This term can compete with the see-saw induced left-handed neutrino masses. However, as the Weinberg operator coupling $\mu^{-1}$ is suppressed by the instanton action, its contribution  is generically subleading relative to the see-saw mass. 

D-brane instantons can also induce dimension-five proton decay  operators \cite{isu07} which  are however sufficiently small for models with large enough string scale.
D-brane instanton generated non-renormalizable operators can also 
serve as an interesting stringy mediation mechanism by connecting the visible and the hidden sector \cite{bf08}.
For an attempt to generate top quark couplings  of $SU(5)$ GUTs via certain non-renormalizable terms, see \cite{CK08}, though such couplings are  usually too small to reproduce a desired top quark mass.

\subsection{Other corrections}
\label{oc}
As discussed in \S\ref{sGenerationFterms} D-brane instantons  can contribute to 
other terms beyond superpotential corrections  such as higher order F-terms and threshold corrections.
While one might naively expect  higher fermionic F-terms to be only  of minor phenomenological interest, it was shown in \cite{AU08} that they become important in the context of moduli stabilization. Once the moduli appearing at a derivative level in the F-terms receive a mass, say the complex structure moduli in the presence of Type IIB 3-form fluxes, they can be integrated out, thus transforming the higher F-term into an effective superpotential at energies below the mass of the moduli.

While our discussion has focused on the generation of otherwise forbidden couplings, instanton corrections can also modify existing physical Yukawa couplings of charged matter fields \cite{ag06}. These corrections can help improve some phenomenological properties of the Standard Model fermion mass matrix. For example, in certain toroidal constructions the mass matrix  has only rank one  at the perturbative level and instanton effects constitute the leading contribution to family mixing \cite{ag06}.

\section{CONCLUSIONS AND OUTLOOK}
\label{sconclusions}

We have reviewed a number  of recent developments  pertinent to
 D-brane instanton effects in ${\cal N}=1$
Type II compactifications to four dimensions. 
This admittedly includes only a small portion
of all the strenuous work on instanton effects in both field and
string theory.  
The  main motivation behind these recent efforts was to
gain a better understanding of the implications of  D-brane instantons 
for string phenomenology. Here the exponentially
suppressed instanton contributions can become
important if by some selection rule the perturbative
contributions are vanishing. These might allow
to solve by a \emph{genuinely stringy mechanism} 
some of the small and large hierarchy problems
present in supersymmetric extensions of the
Standard Model. In addition  instantons
can yield non-trivial contributions to the 
closed string moduli potential and as such 
are essential ingredients for moduli stabilisation, and very consequential for 
cosmic inflation.

We have tried to summarize many recent papers on the development
of a D-brane instanton calculus for the computation of
correlation functions in a D-brane instanton background.
A  rather  coherent picture has emerged for the computation
of correlators corresponding to holomorphic couplings
in the four-dimensional ${\cal N}=1$ supersymmetries
effective action. We have outlined both the (boundary)  conformal field theory
based calculus as well as summarized the rules for local quiver
type models, which are defined on 
highly curved background geometries. All results are in perfect
agreement with each other and with  expectations from
  well-known non-perturbative
dynamics in field theory.

The most obvious types of instantons contributing to a superpotential are of rigid ${\cal O}(1)$ type, but we have described various more complicated instances where extra zero modes can be lifted or saturated in a way compatible with the generation of a  superpotential. We have focused on situations with background fluxes, $U(1)$ instantons on top of single D-branes or configurations with non-trivial instanton-instanton couplings. While the covered results represent the state of the art as of this writing, a deeper understanding of the lifting of fermion zero modes is expected to generalize this picture in the future.

We have also reviewed recent efforts to better understand
the physics of multi-instantons.
Multi-instantons are important to account for
the microscopic behavior of holomorphic four-dimensional
couplings across lines of marginal stability.
Taking the so extracted  multi-instanton calculus
seriously one is  driven to the conclusion that 
the multi-instanton calculus is much richer and more involved
than its field theoretic counterpart.
This can be traced back to the existence of many stringy (exotic) D-brane instantons
in string
theory, which can induce mutual corrections to their instanton
actions. More work is required to complete this picture.
Not unrelatedly, it is also desirable to improve our understanding
of instanton effects on D-terms in the four-dimensional
effective action.

Finally, coming back to our main motivation, we
have summarized some of the implications of D-brane instantons
for phenomenologically important quantities
such as the scalar potential, neutrino masses,
Yukawa couplings and supersymmetry breaking
linear couplings of Polonyi type.
Needless to say that in a top-down approach, 
once a concrete string background
has been fixed, so are  all the non-perturbative
effects. Our list thus only shows which terms
can be generated in principle by what type of instantons; 
it still remains to define concrete models
where instantons with precisely
the right zero mode structure are indeed present. Typically this is particularly sensitive to global constraints such as the tadpole cancellation condition as a hidden sector might introduce extra charged zero modes at its intersections with the instanton.
Another technical challenge is the exact computation and summation of all D-brane instanton effects in a given compactification. While we have outlined some existing technology relating this formidable task to a classical dual via geometric transitions, a long way remains to go until the same degree of sophistication is achieved as in the mirror symmetric computation of worldsheet instanton effects.

\vskip 2cm
\noindent

{\bf Acknowledgements:}

We thank our collaborators M. Aganagic, O. Aharony,  N. Akerblom, R. Argurio, C. Beem, M. Bertolini, B. Florea, S. Franco,  P. Langacker, D. L\"ust, J. McGreevy, S. Moster, E. Plauschinn, R. Richter, N. Saulina, M. Schmidt-Sommerfeld, E. Silverstein and D. Simic for sharing with us the excitement of stringy instantons over the past couple of years. We are indebted to R. Richter for valuable comments on the manuscript and help with the figures.
This work was supported in part by the US Department of Energy under contract DE-AC02-76SF00515, the NSF under
grant PHY-0244728, and the Stanford Institute for Theoretical Physics.

\clearpage \nocite{*}
\bibliography{rev}
\bibliographystyle{utphys}

\end{document}